\documentclass[10pt,twocolumn,letterpaper]{article}

\usepackage{cvpr}              

\usepackage{graphicx}
\usepackage{amsmath}
\usepackage{amssymb}
\usepackage{booktabs}
\usepackage{makecell}
\usepackage{xcolor}
\usepackage{float}
\usepackage[accsupp]{axessibility}

\usepackage[pagebackref,breaklinks,colorlinks]{hyperref}

\usepackage[capitalize]{cleveref}
\crefname{section}{Sec.}{Secs.}
\Crefname{section}{Section}{Sections}
\Crefname{table}{Table}{Tables}
\crefname{table}{Tab.}{Tabs.}

\def\cvprPaperID{10478} 
\def\confName{CVPR}
\def\confYear{2022}

\begin{document}

\title{Multi-Scale Memory-Based Video Deblurring}

\author{Bo Ji \qquad Angela Yao \\
National University of Singapore\\
{\tt\small \{jibo,ayao\}@comp.nus.edu.sg}
\vspace{-0.5em}
}
\maketitle

\begin{abstract}\vspace{-0.85em}
    Video deblurring has achieved remarkable progress thanks to the success of deep neural networks. Most methods solve for the deblurring end-to-end with limited information propagation from the video sequence. 
    However, different frame regions exhibit different characteristics and should be provided with corresponding relevant information.    To achieve fine-grained deblurring, we designed a memory branch to memorize the blurry-sharp feature pairs in the memory bank, thus providing useful information for the blurry query input. To enrich the memory of our memory bank, we further designed a bidirectional recurrency and multi-scale strategy based on the memory bank. Experimental results demonstrate that our model outperforms other state-of-the-art methods while keeping the model complexity and inference time low. The code is available at \href{https://github.com/jibo27/MemDeblur}{https://github.com/jibo27/MemDeblur}.
\end{abstract}
\vspace{-0.3cm}

\section{Introduction}

Video deblurring is a core restoration and enhancement task aiming to recover a sharp video from a blurry input video.  Blur in videos arises from various sources, \eg object motion, camera shake, and depth of field. 
Video deblurring is a highly ill-posed problem as multiple sharp sources may correspond to a single blurred result. The problem becomes especially pertinent as more and more videos are captured by smartphones. 

\begin{figure}
    \centering
    \begin{subfigure}[t]{0.23\textwidth}
        \includegraphics[width=\linewidth]{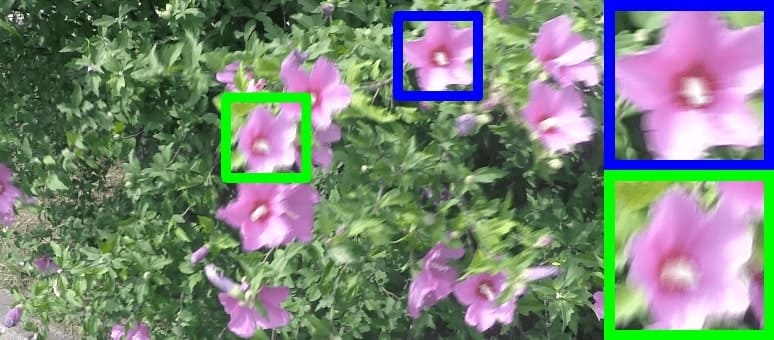}
        \subcaption[]{Frame 0}
    \end{subfigure}
    \begin{subfigure}[t]{0.23\textwidth}
        \includegraphics[width=\linewidth]{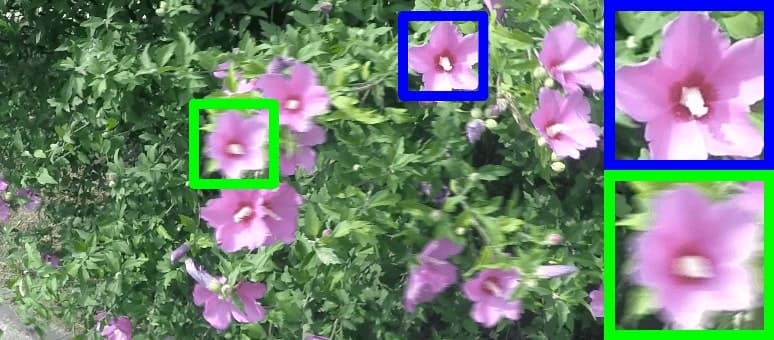}
        \subcaption[]{Frame 16}
    \end{subfigure}
    \begin{subfigure}[t]{0.23\textwidth}
        \includegraphics[width=\linewidth]{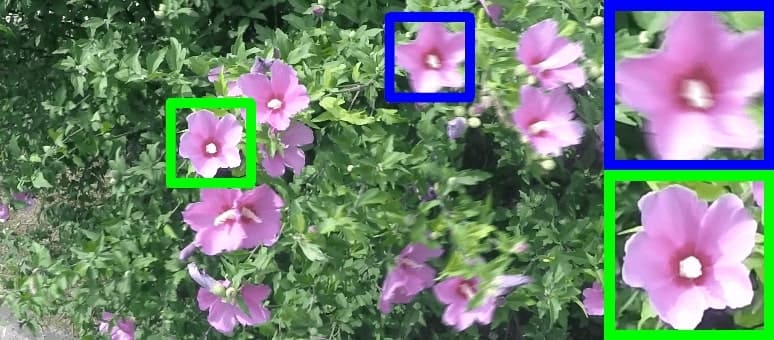}
        \subcaption[]{Frame 20}
    \end{subfigure}
    \begin{subfigure}[t]{0.23\textwidth}
        \includegraphics[width=\linewidth]{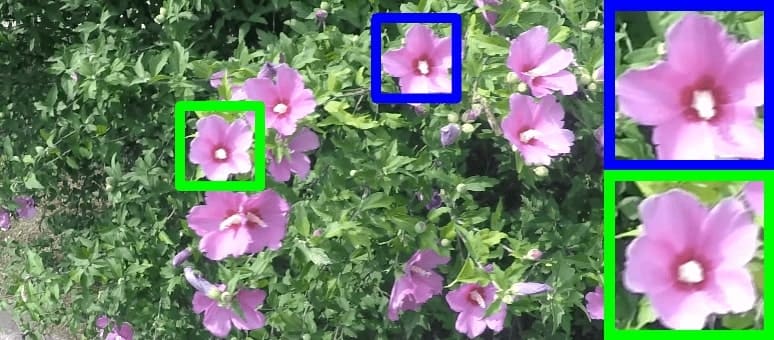}
        \subcaption[]{Frame 86}
    \end{subfigure}
    \caption{Different image regions have different blurry artifacts.\label{fig:problem_setup}
    }
\end{figure}

Video deblurring distinguishes itself from the image deblurring task in that it is critical to use the information from the entire sequence of frames effectively. 
To do so, window-based models~\cite{su2017deep, wang2019edvr,suin2021gated,pan2020cascaded} feed consecutive blurry frames to an encoder-decoder to directly restore a sharp frame. Recurrent models~\cite{nah2019recurrent, zhong2020efficient,wieschollek2017learning,zhou2019spatio,hyun2017online} on the other hand maintain a hidden state across the frames to sequentially propagate features from the first frame to the last. However, these methods cannot utilize the information from \emph{all} the frames in the video sequence. Moreover, neighbouring frames are often used in 
simplistic manners, \eg via an alignment. 

An underexplored aspect in video-deblurring 
is that blurring 
occurs from non-uniform blur kernels. 
The blur artifacts differ not only for the same object across different frames but also for different objects or regions in the same frame. As shown in Fig.~\ref{fig:problem_setup}, the same flower in different frames of a sequence exhibits different extents of blur (compare the blue sample in frame 16 and 86 vs. 0 and 20). Different flowers in the same frame (see frames 16, 20) also have various blur distortions due to object motion. 
This makes spatio-temporal information aggregation difficult as it is necessary to provide the appropriate information to different regions of a frame.

To remedy this problem, we adopt the principle of memory networks~\cite{miller2016key,sukhbaatar2015end,oh2019video,cheng2021rethinking}. Memory networks were originally developed for 
language modeling~\cite{sukhbaatar2015end} 
and are currently popular in vision for video object segmentation~\cite{oh2019video,cheng2021rethinking}. In segmentation, the memory encodes high-level semantics like objects, but they have not been explored in the context of low-level enhancement tasks like deblurring.  The memory networks used in our model record blurry-sharp feature pairs. We compute a spatio-temporal attention between each location of the memory features and that of a query frame region to find 
helpful sharp information. 
The implicit advantage is that even if the blur kernel 
is unknown, we can still extract the corresponding sharp information simply by matching the queried region with memorized features. 

Different from previous works using memory networks, we use memory to supplement information to the deblurring backbone. The core deblurring branch is still responsible for recovering low-level details. 
To enrich the features in the memory bank, we propose a bidirectional and multi-scale structure that better captures information from the entire video sequence at different scales. 
The multi-scale design also allows our model to handle large displacements more effectively.  Our contributions can be listed as follows:\vspace{-0.5em}

\begin{itemize}
    \item We present a novel memory-based architecture 
    that stores blurry and sharp spatio-temporal patterns to achieve fine-grained video deblurring. To the best of our knowledge, we are the first to adopt a memory network for a video enhancement task. 
    \item To increase the diversity of memories in the memory bank, we developed a bidirectional and multi-scale structure based on the memory bank. The multiple scales share the same memory bank, which allows cross-scale matching and effective handling of large motions. 
    \item The experimental results demonstrate that our model achieves superior results than state-of-the-art methods under comparable computational budgets. 
\end{itemize}

\begin{figure}
    \begin{subfigure}[t]{\linewidth}
        \includegraphics[width=\linewidth]{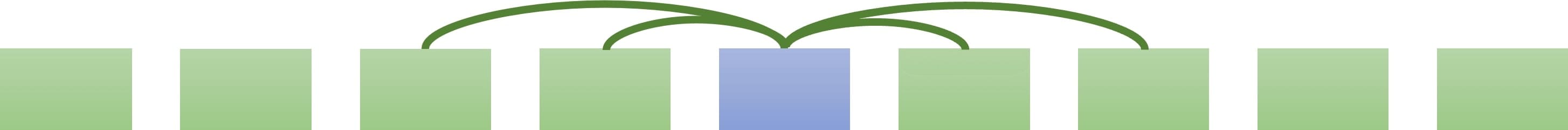}
        \caption{Window-based method~\cite{su2017deep,pan2020cascaded,wang2019edvr}.\label{fig:window_connection}}
    \end{subfigure}
    \begin{subfigure}[t]{\linewidth}
        \includegraphics[width=\linewidth]{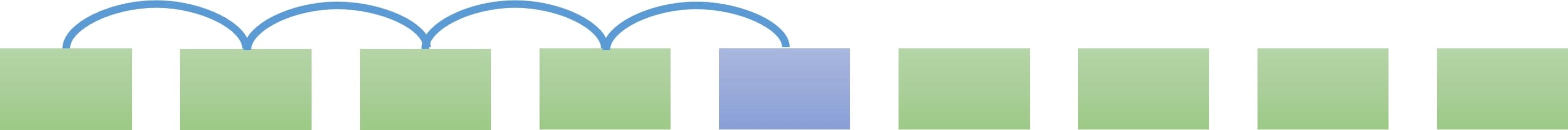}
        \caption{Recurrent method~\cite{nah2019recurrent, wieschollek2017learning,zhou2019spatio,hyun2017online}.\label{fig:recurrent_connection}}
    \end{subfigure}
    \begin{subfigure}[t]{\linewidth}
        \includegraphics[width=\linewidth]{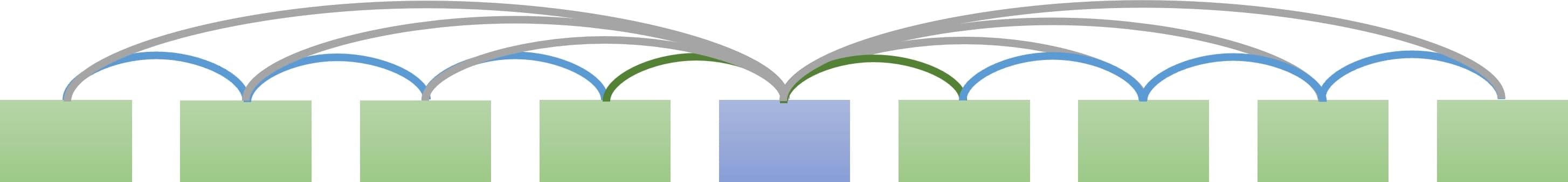}
        \caption{Our proposal.\label{fig:our_connection}}
    \end{subfigure}
    \begin{subfigure}[t]{\linewidth}
        \includegraphics[width=\linewidth]{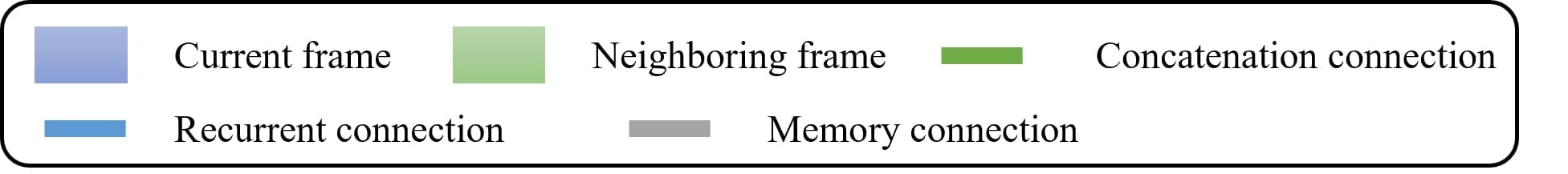}
    \end{subfigure}
    
    \caption{Temporal receptive field comparison between different methods for video deblurring. Window-based approach reconstructs sharp images referencing neighbouring frames within a small fixed-size window.  Recurrent methods collectively aggregate frame features till up-to-date to provide restoration cues. In contrast, our method allows the feature propagation from the entire sequence when restoring the given frame. 
    \label{fig:method_connection}}
    \vspace{-1em}
\end{figure}

\section{Related work}

\paragraph{Video deblurring methods.}
Image or video deblurring methods are widely used in computer vision tasks, such as SLAM~\cite{lee2011simultaneous}, 3D reconstruction~\cite{seok2013dense} and visual tracking~\cite{wu2011blurred}. 
In recent years, deep learning methods have achieved remarkable success in low-level enhancement tasks such as super-resolution~\cite{lim2017enhanced,zhang2018image,sajjadi2018frame}, denoising~\cite{guo2019toward,kim2020transfer} and deblurring~\cite{nah2017deep,tao2018scale, su2017deep}.
Contemporary deep learning video deblurring methods can be roughly divided into window-based and recurrent methods. Window-based methods~\cite{su2017deep,wang2019edvr,pan2020cascaded} (see Fig.~\ref{fig:window_connection}) take 3-7 consecutive frames and try to reconstruct the middle frame. One derivative~\cite{suin2021gated} uses reinforcement learning to select the most effective frames based on the current input and concatenate them with the input of the model. 

Recurrent models~\cite{nah2019recurrent, zhong2020efficient,wieschollek2017learning,zhou2019spatio,hyun2017online} deblur frames in sequence, maintaining a recurrent hidden state to propagate information from frame to frame (see Fig.~\ref{fig:recurrent_connection}). ESTRNN~\cite{zhong2020efficient} proposes a spatio-temporal attention module to fuse the neighboring features further. Our model is also recurrent, though we supplement the hidden vector with information from the memory network to extend the range of contextual information (see Fig.~\ref{fig:our_connection})

Existing methods, be it window-based or recurrent, do not consider the characteristics of a localized region within a frame.  Furthermore, they utilize only a portion of the frames in the sequence. Our method uses memory banks to provide appropriate spatio-temporal information for each region and exploits the features of all frames.

\vspace{-0.4em}\paragraph{Memory networks.}
Memory networks are a class of learning models~\cite{weston2014memory} that construct an external \emph{memory bank} module to store potentially useful features for future use. Memory networks were originally developed for NLP applications~\cite{weston2014memory,sukhbaatar2015end,miller2016key}, but the property of storing features lends itself naturally to being used in video where there is large redundancy across frames.  As such, memory networks have been extended to several other video tasks such as movie understanding~\cite{na2017read}, object tracking~\cite{yang2018learning} and, more recently, video object segmentation~\cite{oh2019video,lu2020video,cheng2021rethinking,seong2020kernelized}. 

To the best of our knowledge, memory networks have not been used for low-level enhancement tasks like video deblurring. As memory networks traditionally encode semantic information, it is not known if they can effectively handle the underlying feature reconstruction of low-level tasks. Different from existing works, we designed a new memory network customized for the video deblurring task. Especially noteworthy is our novel multi-scale memories which can be shared across different scales.

\section{Approach}\label{sec:approach}

\begin{figure*}
    \centering
    \includegraphics[width=\linewidth]{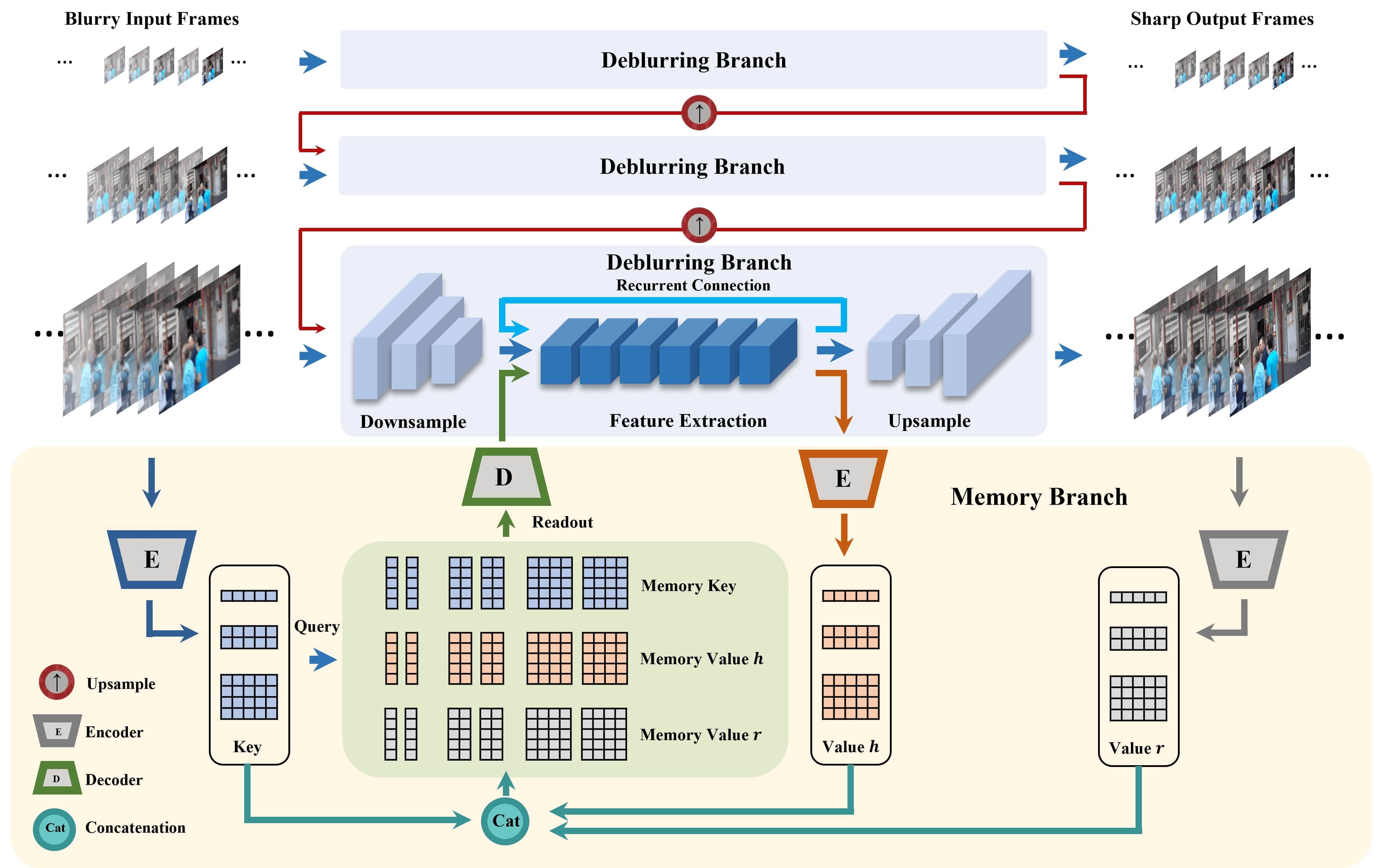}
    \vspace{-1.4em}
    \caption{\textbf{Overview.} The input is represented at three different scales. We recover the entire video sequence at the corresponding scale, starting from the lowest scale, until the sequence of the original input size is restored. For each scale, we sequentially restore from the first to the last frame. The same memory bank is shared between scales, which ensures information propagation across multiple scales.  \label{fig:approach}}
    \vspace{-1em}
\end{figure*}

We adopt a multi-scale memory-based structure to solve the video deblurring task. There are two branches: the deblurring branch and the memory branch (see Fig.~\ref{fig:approach}).  The deblurring branch features a bidirectional recurrent structure, while the memory branch stores blurry-sharp feature pairs. 

\subsection{Preliminaries}

Given a blurry video sequence $\mathbf{I}\!=\!\{I_1, \dots, I_i, \dots, I_{N}\}$ as input, our objective is to recover a deblurred output sequence $\mathbf{R} = \{R_1, \dots, R_i, \dots, R_{N}\}$. Considering a blurry frame $i$, the recovered frame $R_i$ can be given by a learned decoding or ``upsampling''\footnote{We abuse the terms ``upsampling'' and ``downsampling'' to emphasize a transformation in both features space and spatial dimensionality.} $\mathcal{U}(\cdot)$ of a hidden vector $h_i$, \ie 
\begin{align}
\begin{split}
    R_i & = \mathcal{U}(h_i), 
    \\
    \text{where} \;\; h_i & = \mathcal{F}([x_i, x_{i-1}, h_{i-1}, m_i]), 
\end{split}\label{eq:R_i_and_h_i}
\end{align}
\noindent 
where $\mathcal{F}$ represents a recurrent feature extraction module, $\mathcal{D}$ is a learned encoding or ``downsampling'', $x_i\!=\!\mathcal{D}(I_i)$, $x_{i-1} = \mathcal{D}(I_{i-1})$ and $m_i$ is a retrieved deblurred memory feature. From~\cref{eq:R_i_and_h_i}, one can see that the feature extraction module $\mathcal{F}$ has a recurrent structure, in that $h_i$ relies on the current downsampled input feature $x_i$, retrieved $m_i$ and also the previous $h_{i-1}$. We also feed the previous downsampled feature $x_{i-1}$ into the feature extraction module as this has been shown to be effective~\cite{gast2019deep}.

\subsection{Memory-enhanced feature aggregation}\label{sec:memory}

The memory bank stores the blurry-sharp feature pairs in the latent space so that the relevant sharp features are retrieved with the blurry query input. We consider past restored frames as memory frames saved in the memory bank and the current blurry frame as the query frame. 

Saved memory frames follow a key-value format. The key is encoded from the blurry frame $I_i$, while the values are features extracted from the corresponding hidden features $h_i$ and deblurred result $R_i$. Given a query frame, we compare it with memory frames in the key space and retrieve the associated values. Then, we decode the values back as relevant effective information for the query input. For clarity, we ignore the bidirectional design and scale level $s$ when describing the details of the memory alignment.

\paragraph{Key encoding.}
We reuse the downsampling module $\mathcal{D}$ to reduce input frame $I_i$ into $x_i$, though the spatial size is still too large for the matching between the query and memory keys. Inspired by ~\cite{oh2019video}, we use a key encoder $\mathcal{K}$ to further reduce the computational overhead to a key $k_i$: 
\begin{align}
    k_i = \mathcal{K}(x_i).
\end{align}

\paragraph{Value encoding.} 
Regarding the value, we apply the encoder-decoder architecture and save the values in the latent space to reduce the memory cost and computational cost. The value encoding module $\mathcal{V}$ is responsible for encoding the corresponding pair into their latent values:
\begin{align}
\begin{split}
    v_i^r & = \mathcal{V}_r([x_i, r_i]), \\
    v_i^h & = \mathcal{V}_h([x_i, h_i]),
\end{split}
\end{align}
\noindent where $r_i = \mathcal{D}(R_i)$ and  $[\cdot]$ represents the concatenation operation. While it is feasible to directly store a value triplet, \ie $v_i = \mathcal{V}([x_i, r_i, h_i])$, we find through preliminary studies that such a design is too limiting and degrades the performance (see Section~\ref{sec:network_analysis}), likely because it does not distinguish the different roles of $r_i$ and $h_i$.

For more flexibility, we store $v_i^r$ and $v_i^h$ in two different memories.  For storage, we concatenate them with the current memory $\mathbf{v}^{r,M}$, $\mathbf{v}^{h, M}$ and update:
\begin{align}
\begin{split}
    \mathbf{k}^M & = [\mathbf{k}^M, k_i] \\
    \mathbf{v}_i^{r, M} & = [\mathbf{v}_{i-1}^{r, M}, v_i] \\
    \mathbf{v}_i^{h, M} & = [\mathbf{v}_{i-1}^{h, M}, v_i].
\end{split}
\end{align}

Note that the two values share the same key. While it is possible to update $v^M$ with values at every temporal index $i$, increasing the frequency of memorization increases the memory cost without necessarily improving the performance. As such, we update the memory with every $T$ frames. 

\paragraph{Memory readout.} 
Here, we ignore the frame index $i$ for a given query key $k_i$ and use $q$ to denote the $q$-th location of $k_i$ as $k_q^Q$. Similarly, we denote $k_p^M$ as the $p$-th location of the memory key $\mathbf{k}^M$. Then, we compute the affinity matrix $\mathbf{S}$ between $k_q^Q$ and  $k_p^M$:
\begin{align}
    \mathbf{S}_{q, p} = d(k_q^Q, k_p^M),
\end{align}
where $d$ is any similarity measure. The affinity matrix $\mathbf{S}$ is then normalized by a softmax to matrix $\mathbf{W}$:
\begin{align}\label{eq:calc_w}
    \mathbf{W}_{q, p} = \frac{\exp({\mathbf{S}_{q, p}})}{  \sqrt{C^k} \cdot \sum_z{ \exp({\mathbf{S}_{q, z}})  } },
\end{align}
where $C^k$ is the key dimension and $\sqrt{C^k}$ serves as a normalization term~\cite{vaswani2017attention}. This operation is a form of spatio-temporal attention, where $\mathbf{W}$ can be viewed as an attention map, as we are looking for locations on the memory frames that are most relevant for restoring the current query location.

The matrix $\mathbf{W}$ is shared across the readout operation for two memory values, $v_i^r$ and $v_i^h$, where $i$ is the frame index. The readout memory for the query key $k_i$ is given as
\begin{align}
\begin{split}
    v_i^{r, Q} & = \mathbf{v}_i^{r, M} \mathbf{W} \\
    v_i^{h, Q} & = \mathbf{v}_i^{h, M} \mathbf{W},
\end{split}
\end{align}
where $\mathbf{v}_i^{r, M}$ and $\mathbf{v}_i^{h, M}$ represent the current memory for the two pairs when we try to restore $i$-th frame $I_i$, $v_i^{r, Q}$ and $v_i^{h, Q}$ are the corresponding aggregated memory outputs.

\paragraph{Memory decoding.} The readout memory $v_i^{r, Q}$ and $v_i^{h, Q}$ are further decoded by the corresponding decoders $\mathcal{G}_r$ and $\mathcal{G}_h$:
\begin{align}
\begin{split}
    m_i^{r, Q} & = \mathcal{G}_r(v_i^{r, Q}) \\
    m_i^{h, Q} & = \mathcal{G}_h(v_i^{h, Q}).
\end{split}
\end{align}
We then concatenate the two decoded memories to get $m_i$:
\begin{align}
    m_i = [m_i^{r, Q}, m_i^{h, Q}],
\end{align}
where $m_i$ is the final memory output for the input frame $I_i$. As the encoders $\mathcal{K}$ and $\mathcal{V}$ downsample the input feature $x_i$, the decoder $\mathcal{G}$ upscales the memory back to the original spatial size.

\subsection{Bidirectional recurrency}\label{sec:bidirection}
If a video sequence is deblurred in a uni-directional manner, \eg forwards, then frame $I_i$ receives context and memory only from frames up to $i-1$. Yet, there likely exist helpful details in frames after $i+1$. Therefore, we make the recurrent framework bidirectional, incorporating a forward and a backward pass of the sequence to further enrich the memory bank. 
Just as we establish a memory bank in the forward pass of the sequence, we can also do the same for the backward pass.
Note that the sharp frame is restored with information from both passes. As we perform the backward pass first, there are no sharp frames available during the backward procedure, so we encode only the hidden features $h_i$ when establishing the backward memory bank. 

For the backward pass, the input sequence is fed in as $\{I_N, I_{N-1}, \dots, I_1\}$, where each recurrent unit operates as
\begin{align}
    h_i^b = \mathcal{F}_b([x_i, x_{i + 1}, h_{i+1}^b, m_i^b]),
\end{align}
and $h_i^b$ is the hidden state of the backward module $\mathcal{F}_b$ for $i$-th downsampled feature $x_i$, $m_i^b$ is the aggregated memory with the query key $k_i$. We store the forward and backward features separately in two memory banks.  This allows the two memories to be extracted separately during the forward pass, which we find to be  more beneficial for the model to distinguish and learn.
Then, in the forward pass, the feature extraction module can utilize the backward memory to refine the current features:
\begin{align}\label{eq:backward}
    h_i^f = \mathcal{F}_f([x_i, x_{i - 1}, h_{i-1}^f, m_i^f, m_i^b]),
\end{align}
where $h_i^f$ is the hidden state of the forward module $\mathcal{F}_f$, and $m_i^f$ is the memory aggregated from the previous frames.  We fuse $h_i^f$ and $h_i^b$ using one convolutional layer to get $h_i$:
\begin{align}\label{eq:fusion}
    h_i = \text{conv}([h_i^f, h_i^b])
\end{align}
\noindent The final reconstruction operation is given by Eq.~\ref{eq:R_i_and_h_i}.

The bidirectional design is important. The backward module not only provides the visibility of the future frames, but also gives more sufficient information for the restoration of the first frame $x_1$ in the forward pass. In Eq.~\ref{eq:backward}, as $x_1$ is the first frame in the video sequence, $x_{i-1}$, $h_{i-1}^f$ and $m_i^f$ are all initialized as zeros. If we do not retrieve memory from the backward pass, i.e., $m_i^b$, the model only uses the information of the current frame to perform the deblurring. In that case, it downgrades into a single-image deblurring task and would produce an inferior result. Moreover, in Eq.~\ref{eq:fusion}, the backward pass provides an additional helpful feature $h_i^b$.

\subsection{Multi-scale design}
Memorizing the relationship among the downsampled feature $x_i$, hidden state $h_i$, and sharp frame $R_i$ occupies memory and computational cost. To balance efficiency and performance, we only memorize every $T$ frame, as mentioned in Section~\ref{sec:memory}, and perform the temporally intensive memorization for downsample sequences. This motivates us to adopt a multi-scale strategy.  Multi-scaling has proven to be effective in image deblurring~\cite{nah2017deep,tao2018scale} but is not yet explored in video tasks. 

Adopting a multi-scale strategy causes limited memory and computational overhead, but comes with the key advantage of allowing the matching between patterns at different scales. It also handles large displacement more  effectively, as downscaling reduces the size of the original (large)  displacements. 
To consider multiple scales, 
we downsample the input frame $I_i\!\in\!\mathbb{R}^{H\times W\times C}$ into $I_i^2\!\in\!\mathbb{R}^{H/2\times W/2 \times C}$ and $I_i^3\!\in\!\mathbb{R}^{H/4\times W/4 \times C}$, where the original input $I_i$ can be considered as $I_i^1$. Note that the downsampling and upsampling in the multi-scale strategy are different from downsampling module $\mathcal{D}$ and upsample module $\mathcal{U}$ in the deblurring branch. The former uses a deterministic method, \eg bilinear interpolation, as the objective is purely scaling and not feature extraction. 
At scale level $s$ for the forward pass, we obtain the hidden state by
\begin{align}\label{eq:multi_scale_forward}
    h_i^{f, s} = \mathcal{F}_f([x_i^s, x_{i-1}^s, x_i^{s+1 \uparrow}, m_i^{f,s}, m_i^{b,s}]),
\end{align}
where $h_i^{f, s}$ is the hidden state at scale level $s$, $x_i^{s+1 \uparrow}$ is the upscaled version of the downsampled feature from the deblurred frames at scale $s\!+\!1$, \ie $x_i^{s+1 \uparrow}\!=\!(\mathcal{D}(R_i^{s+1}))\uparrow$, $m_i^{f,s}$ and $m_i^{b,s}$ are aggregated memories from forward and backward memory banks for the given input $x_i^s$. The $\uparrow$ denotes bilinear upsampling. 

The forward module of different scales share the same memory bank. This makes it possible for the query frame to match features to multiple scales and enhances the memory utilization.  The shared-scale memory is a key distinction of our work from previous multi-scale methods.

\subsection{Architecture}
We use residual dense blocks~\cite{zhang2018residual}, residual blocks~\cite{lim2017enhanced} and transposed convolution as the basic building blocks for $\mathcal{D}$, $\mathcal{F}$ and $\mathcal{U}$ in the deblurring branch, respectively. For our memory branch, the $3$ encoders $\mathcal{K}$, $\mathcal{V}_r$ and $\mathcal{V}_h$ are of the same architecture, which is residual blocks followed the first stage of pre-trained ResNet50~\cite{he2016deep}. The decoder $\mathcal{G}$ combines the output of $2$ decoders $\mathcal{G}_r$ and $\mathcal{G}_h$, which contains residual blocks and a single $\times 4$ pixel shuffle layer~\cite{shi2016real}. More details are in the Supplementary.

\section{Experiments}\label{sec:experiments}

\subsection{Setting}\label{sec:exp_setting}
\textbf{Dataset.} We experimented on the GOPRO dataset~\cite{nah2017deep}, which features $22$ training sequences with $2103$ frames and $11$ validation sequences with $1111$ frames.  We experimented with two variants:
the original version~\cite{pan2020cascaded} and the 
downsampled version with gamma correction~\cite{nah2019recurrent,zhong2020efficient}, which downsamples the original videos from $1280\times 720$ to $960\times 540$ to reduce noise and video compression artifacts. 

\textbf{Evaluation metrics.} We use the peak signal-to-noise ratio (PSNR) and SSIM~\cite{wang2004image} to evaluate the deblurred results. For complexity, we compare the runtime and the multiply-accumulate operation (MAC).
The runtime is calculated per frame on the input video containing $100$ frames using a single NVIDIA RTX A5000 GPU. Both runtime and MAC assume the input frame is of shape $720\times 1280\times 3$.

\textbf{Implementation details.} For training, we used the ADAM optimizer~\cite{kingma2014adam} with parameters $\beta_1\!=\!0.9$, $\beta_2\!=\!0.999$ and $\epsilon\!=\!10^{-8}$. The initial learning rate was set as $0.0005$ and was decayed by half in $[200, 350, 450, 500]$ epochs. We trained the model for $600$ epochs in total. For augmentation, we applied random rotations and flipping. To manage the memory for scale $s$, we set $T_1\!=\!5$, $T_2\!=\!2$ and $T_3\!=\!1$. The training patch size was $256\times 256$, and the batch size was $8$. The training subsequence contained $8$ frames.
To manage the memory cost and run-time, 
we empirically maintained a maximum of five recent frames in the memory.
This technique reduced the runtime to one-fifth of the original
, while the PSNR dropped by only $0.02$dB. 
We borrowed the multi-scale content loss~\cite{nah2017deep} to train our model, but we replaced the mean-square error with Charbonnier loss~\cite{charbonnier1994two}.

\subsection{Comparisons with state of the art}

\begin{table*}
    \caption{Quantitative comparison on the downsampled GOPRO dataset~\cite{nah2017deep}. \label{table:sota_downsampled_gopro}}
    \vspace{-0.1cm}
    \centering
    \resizebox{\textwidth}{!}{
    \def\arraystretch{1.1}
    \begin{tabular}{@{\extracolsep{4pt}}ccccccccccc @{}}
        \Xhline{3\arrayrulewidth}
        Model & STRCNN~\cite{hyun2017online} & DBN~\cite{su2017deep} & IFIRNN($c2h2$)~\cite{nah2019recurrent} & IFIRNN($c2h3$)  & ESTRNN($B_9C_{80}$)~\cite{zhong2020efficient} & ESTRNN($B_9C_{90}$) & Ours (Slim) & Ours\\
        \hline
        PSNR & 28.74 & 29.91 & 29.92 & 29.97 & 30.79 & 31.07 & 31.21 & 31.77  \\
        SSIM & 0.8465 & 0.8823 & 0.8838 & 0.8859 & 0.9016 & 0.9023 & 0.9203 & 0.9275 \\
        GMACs & 276.2 & 784.75 & 167.09 & 217.89 & 163.61 & 206.70 & 197.34 & 344.49  \\
        \Xhline{3\arrayrulewidth}
    \end{tabular}
    }
\end{table*}

\begin{table*}
    \caption{Quantitative comparison on the original GOPRO dataset~\cite{nah2017deep}. We do not fill in the GMACs of STFAN~\cite{zhou2019spatio} as the GMACs calculated for filter adaptive layer may be not accurate. \label{table:sota_original_gopro}}
    \vspace{-0.1cm}
    \centering
    \resizebox{\textwidth}{!}{
    \def\arraystretch{1.1}
    \begin{tabular}{@{\extracolsep{4pt}}cccccccccc @{}}
        \Xhline{3\arrayrulewidth}
        Model & SRN~\cite{tao2018scale} & DBN~\cite{su2017deep}  & Kim et al.~\cite{hyun2017online} & EDVR~\cite{wang2019edvr} & STFAN~\cite{zhou2019spatio} & ESTRNN~\cite{zhong2020efficient} & CDVD-TSP~\cite{pan2020cascaded} &  Ours\\
        \hline
        PSNR & 30.29 & 27.31 & 26.82 & 26.83 & 28.59 & 30.91 & 31.67 & 31.76 \\
        SSIM & 0.9014 & 0.8255 & 0.8245 & 0.8426 & 0.8426 & 0.9091 & 0.9275 & 0.9230\\
        GMACs & 1527.01 & 784.75  & 276.2 & 468.25 & - & 204.19 & 5122.29 & 344.49  \\
        \Xhline{3\arrayrulewidth}
    \end{tabular}
    }
    \vspace{-1em}
\end{table*}

\begin{table}
    \caption{Runtime comparison with state-of-the-art models.   \label{table:sota_runtime}}
    \vspace{-0.2cm}
    \centering
    \resizebox{\columnwidth}{!}{
    \def\arraystretch{1.1}
    \begin{tabular}{@{\extracolsep{4pt}}ccccc @{}}
        \Xhline{3\arrayrulewidth}
        Model & SRN & STRCNN & DBN & IFI-RNN \\
        \hline
        Runtime (s) & 0.222 & 0.0448 & 0.085 & 0.054 \\
        \hline
        Model & ESTRNN & CDVD-TSP & Ours(Slim) & Ours \\
        \hline
        Runtime (s) & 0.083 & 1.015 & 0.079 & 0.191 \\
        \Xhline{3\arrayrulewidth}
    \end{tabular}
    }
    \vspace{-1.5em}
\end{table} 

\begin{figure*}
    \centering
    \begin{subfigure}[t]{0.16\textwidth}
        \includegraphics[width=\linewidth]{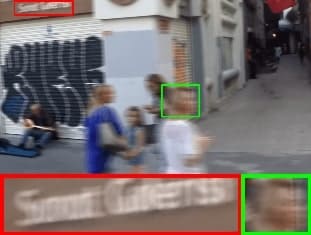}
        \subcaption{Input}
    \end{subfigure}
    \begin{subfigure}[t]{0.16\textwidth}
        \includegraphics[width=\linewidth]{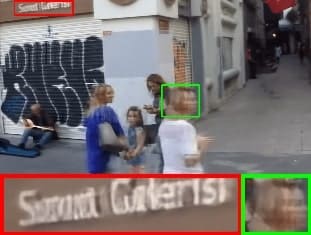}
        \subcaption{DBN~\cite{su2017deep}}
    \end{subfigure}
    \begin{subfigure}[t]{0.16\textwidth}
        \includegraphics[width=\linewidth]{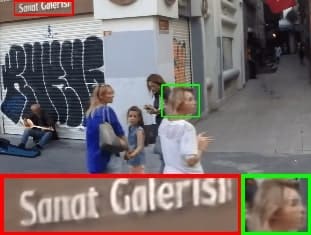}
        \subcaption{ESTRNN~\cite{zhong2020efficient}}
    \end{subfigure}
    \begin{subfigure}[t]{0.16\textwidth}
        \includegraphics[width=\linewidth]{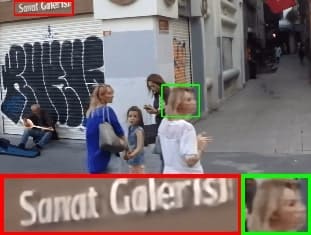}
        \subcaption{CDVD-TSP~\cite{pan2020cascaded}}
    \end{subfigure}
    \begin{subfigure}[t]{0.16\textwidth}
        \includegraphics[width=\linewidth]{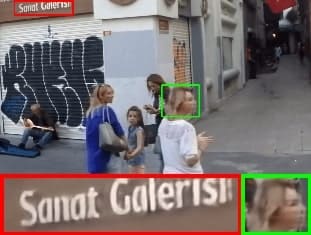}
        \subcaption{Ours}
    \end{subfigure}
    \begin{subfigure}[t]{0.16\textwidth}
        \includegraphics[width=\linewidth]{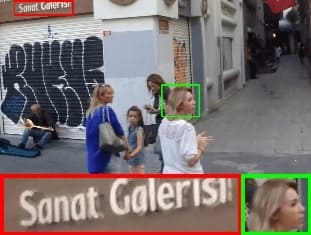}
        \subcaption{Ground-truth}
    \end{subfigure}
    \vspace{-0.1cm}
    \caption{Qualitative comparisons on the original GOPRO dataset~\cite{nah2017deep}. \label{fig:qualitative_gopro}}
    \vspace{-0.1em}
\end{figure*}

\begin{figure*}
    \centering
    \begin{subfigure}[t]{0.48\textwidth}
        \includegraphics[width=0.32\linewidth]{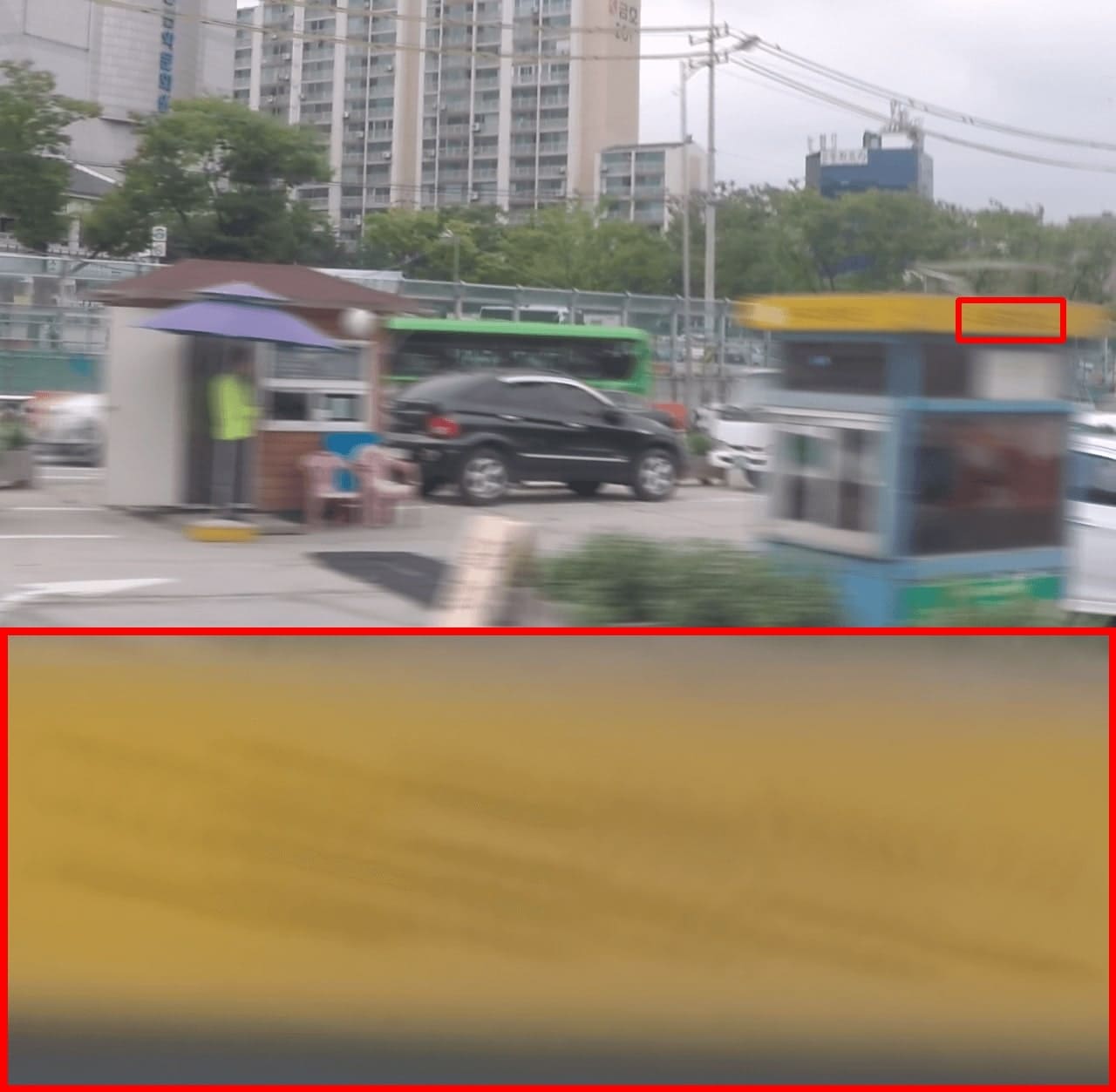}
        \includegraphics[width=0.32\linewidth]{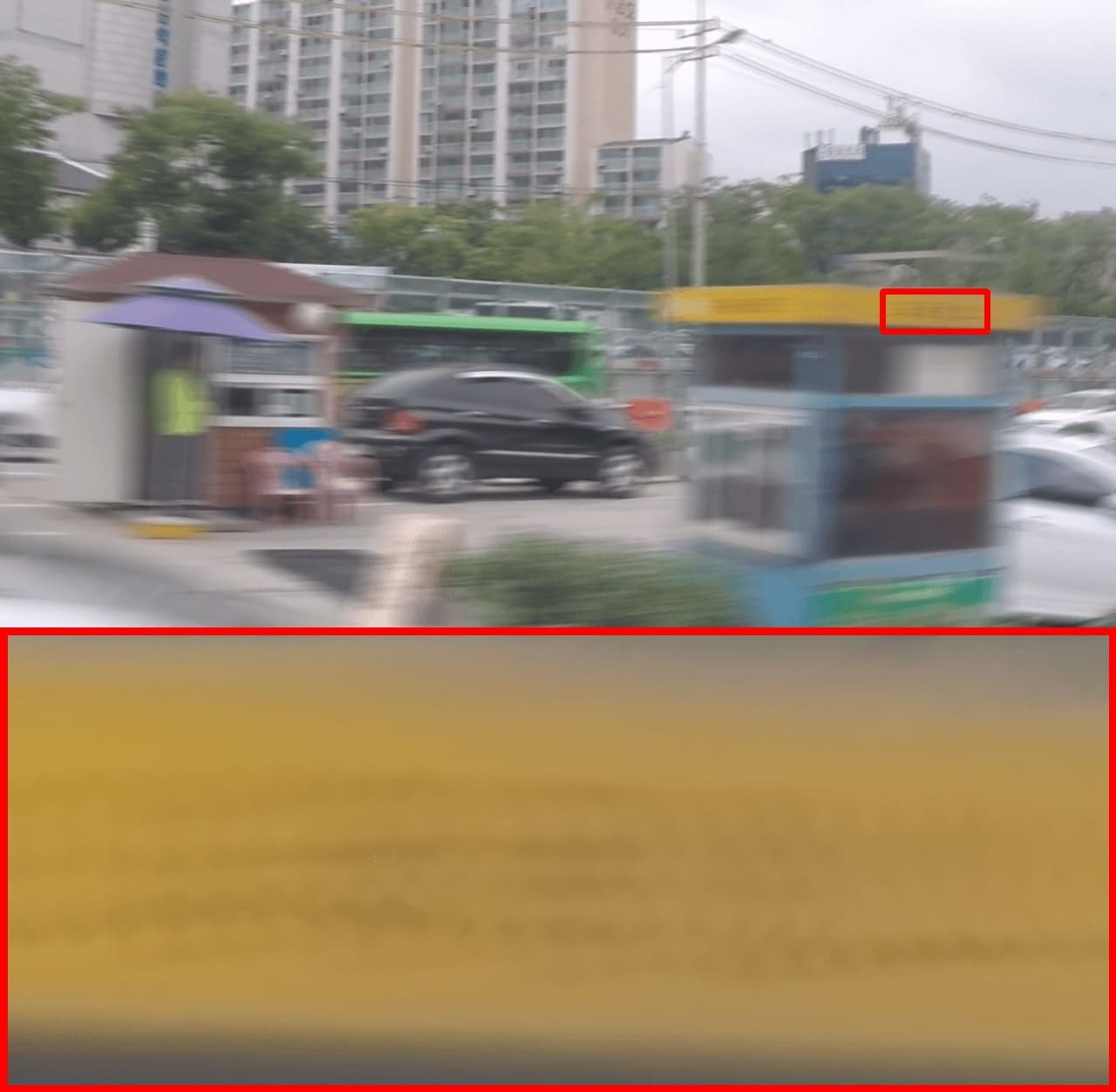}
        \includegraphics[width=0.32\linewidth]{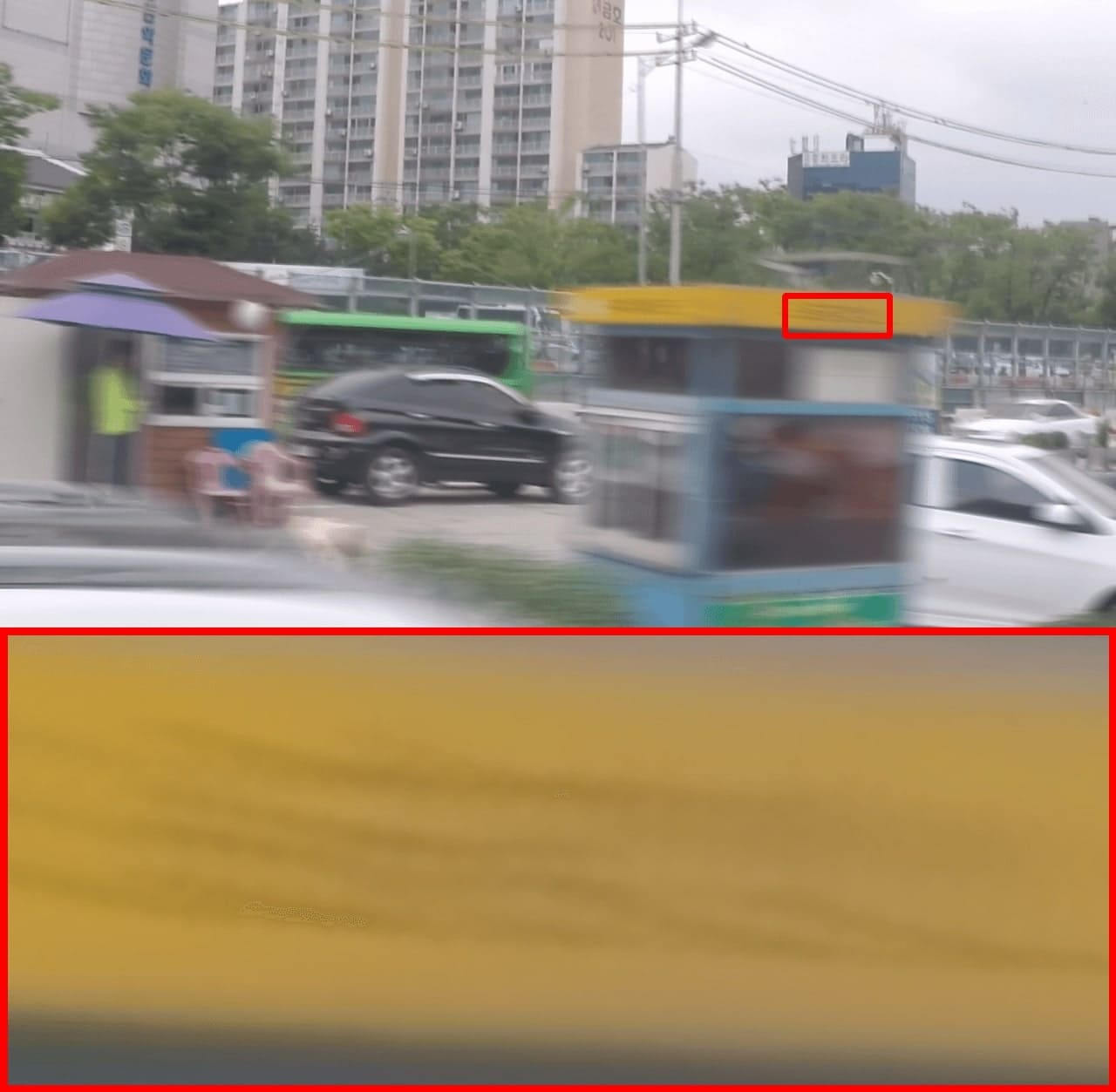}
        \subcaption{Input}
    \end{subfigure}
    \begin{subfigure}[t]{0.48\textwidth}
        \includegraphics[width=0.32\linewidth]{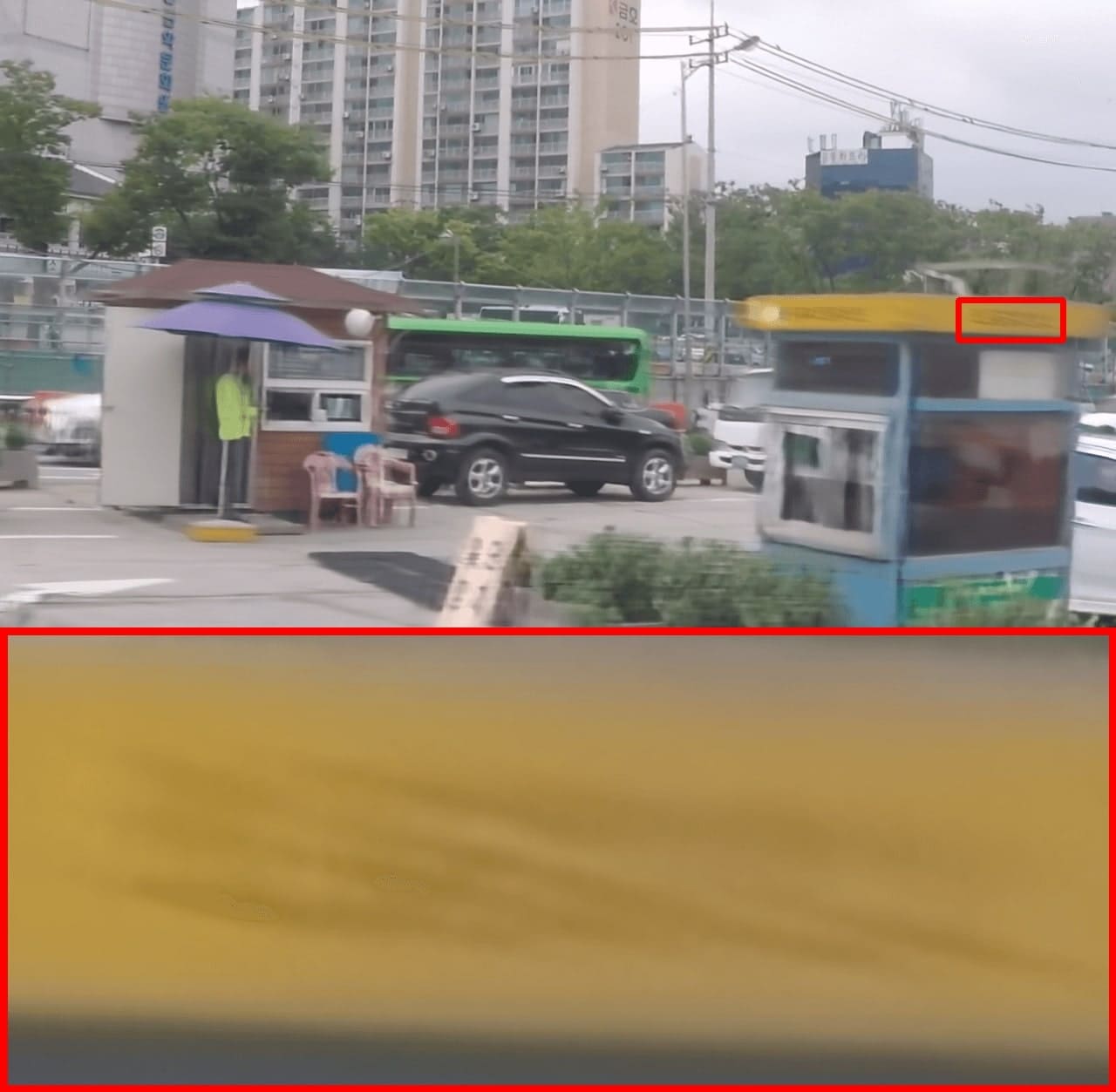}
        \includegraphics[width=0.32\linewidth]{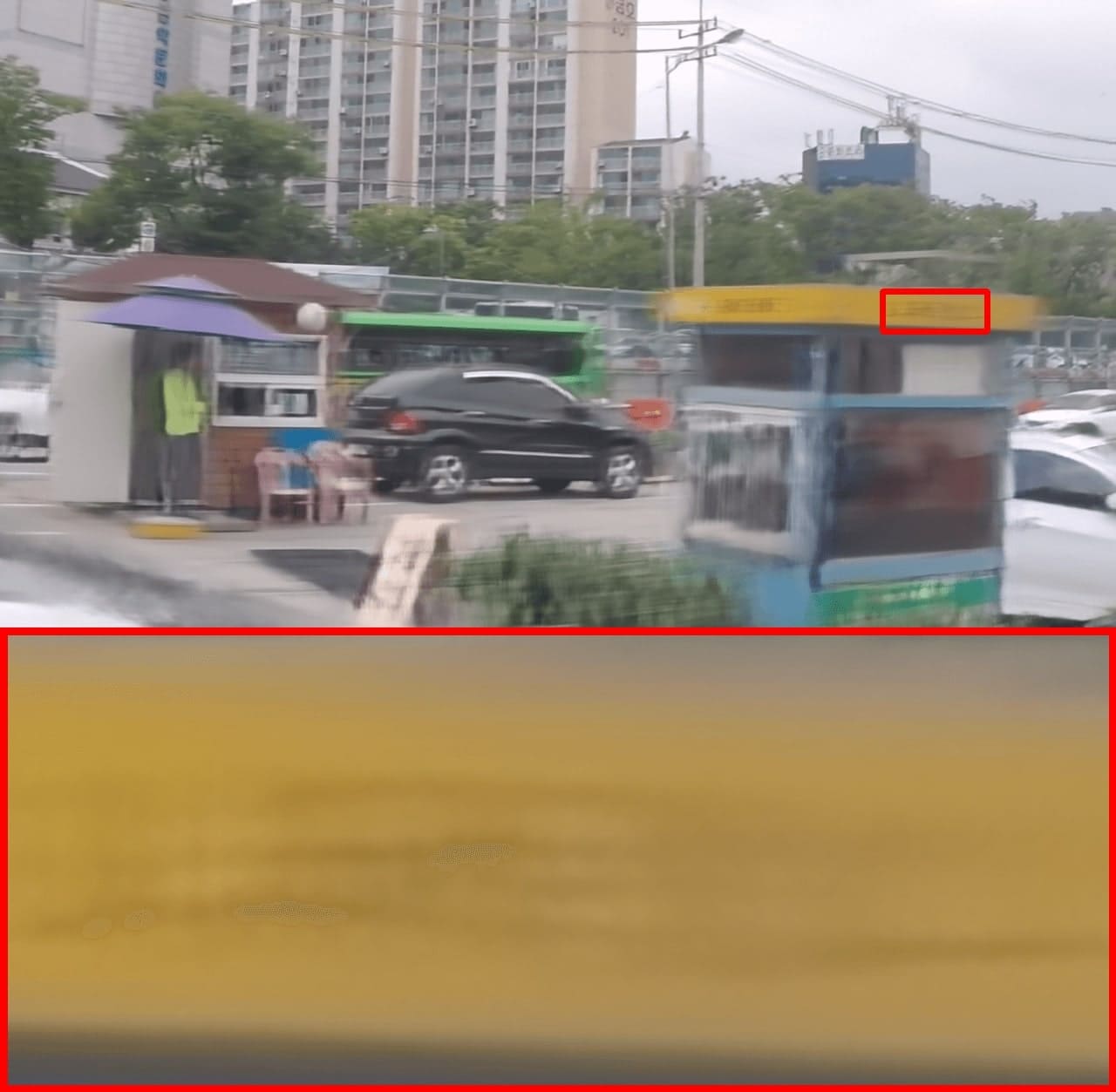}
        \includegraphics[width=0.32\linewidth]{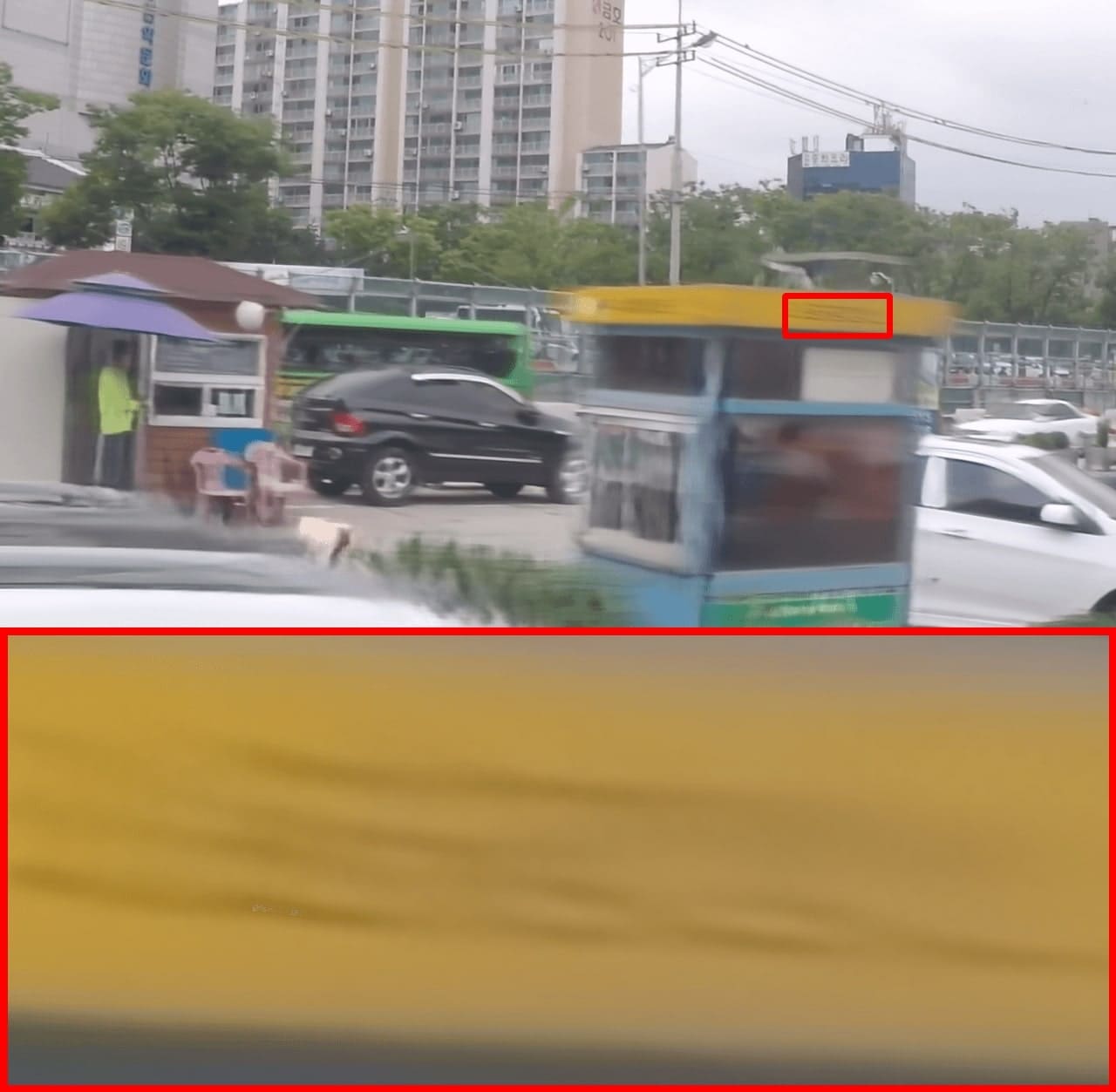}
        \subcaption{DBN~\cite{su2017deep}}
    \end{subfigure}
    \begin{subfigure}[t]{0.48\textwidth}
        \includegraphics[width=0.32\linewidth]{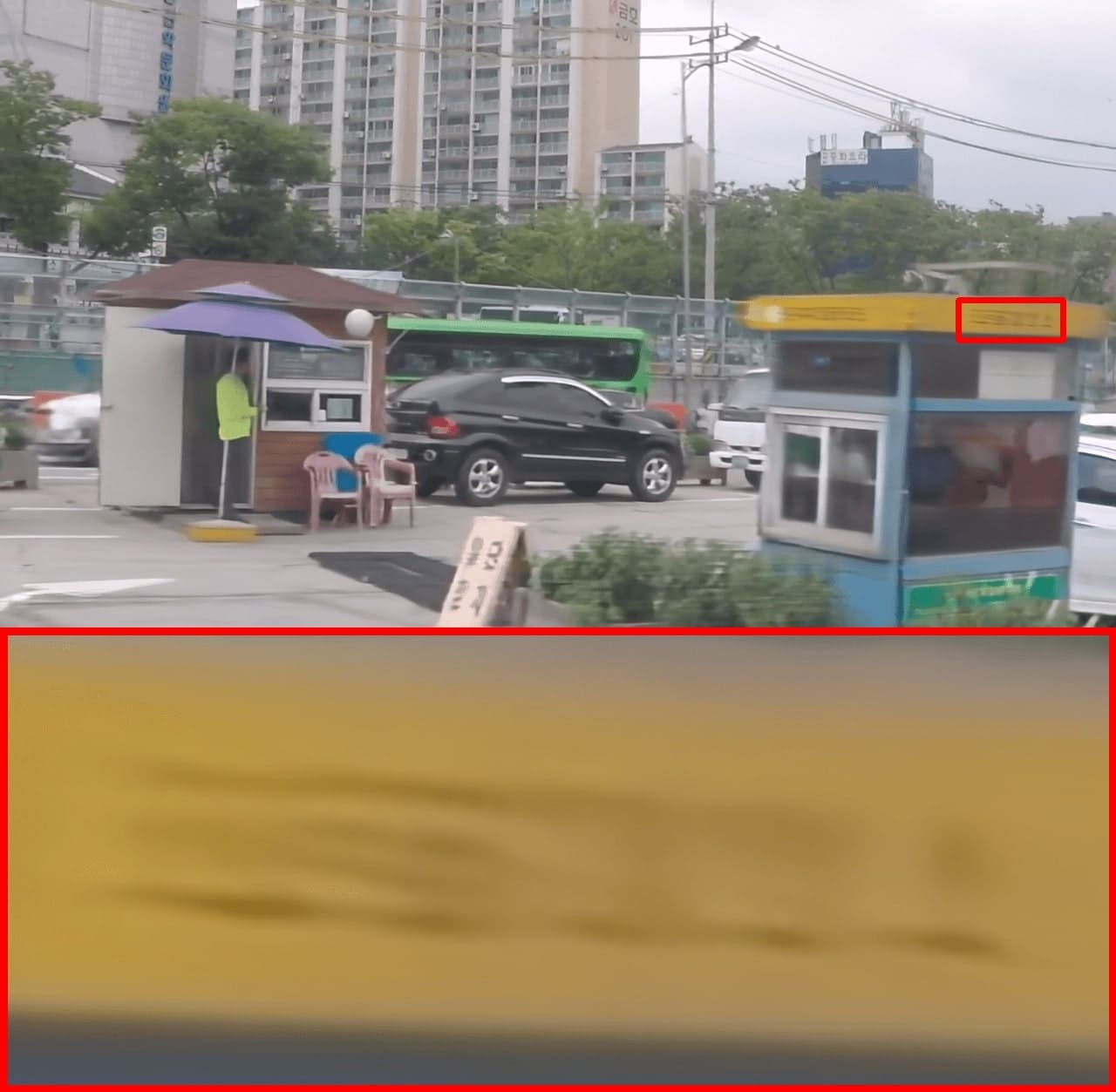}
        \includegraphics[width=0.32\linewidth]{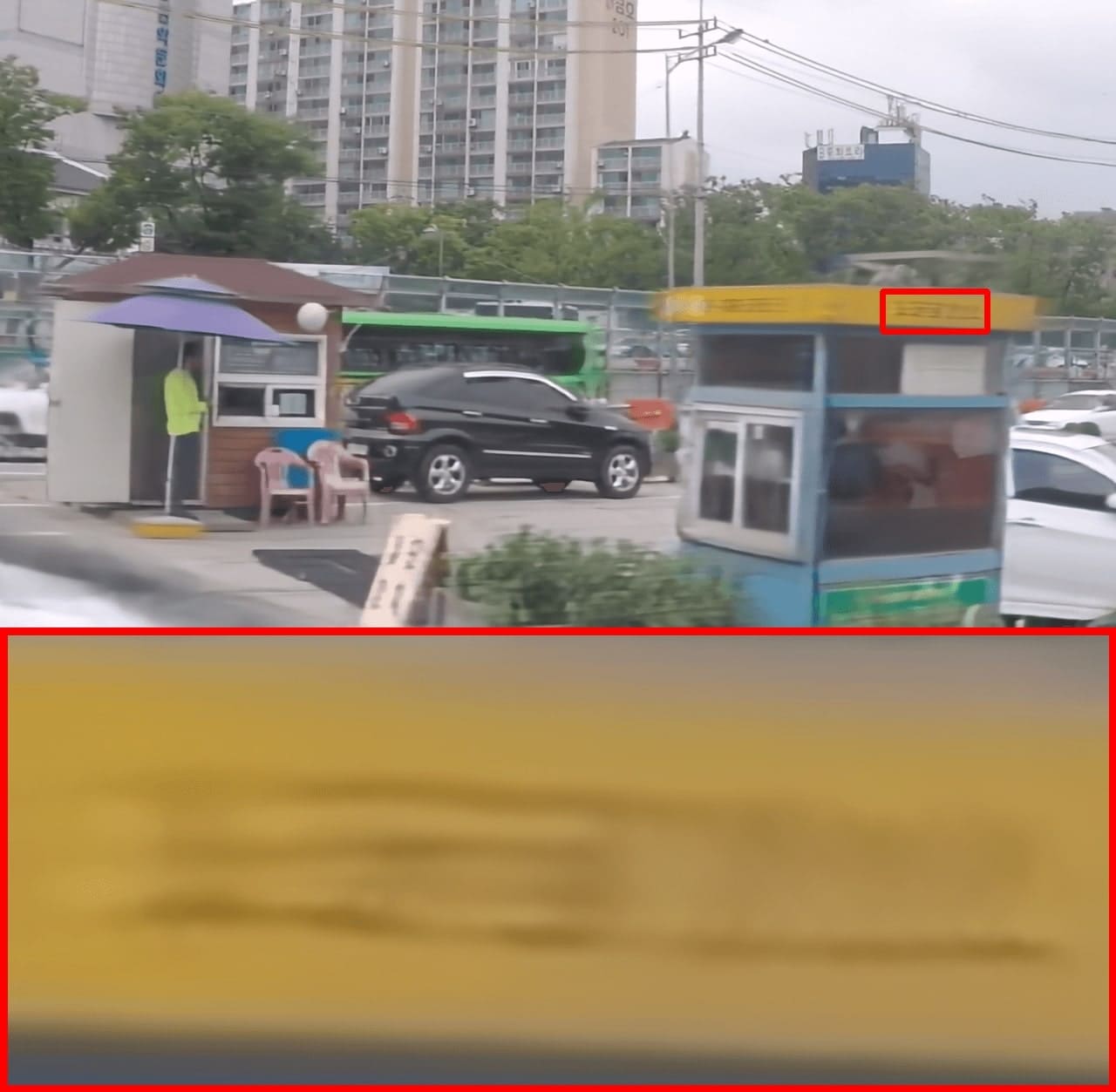}
        \includegraphics[width=0.32\linewidth]{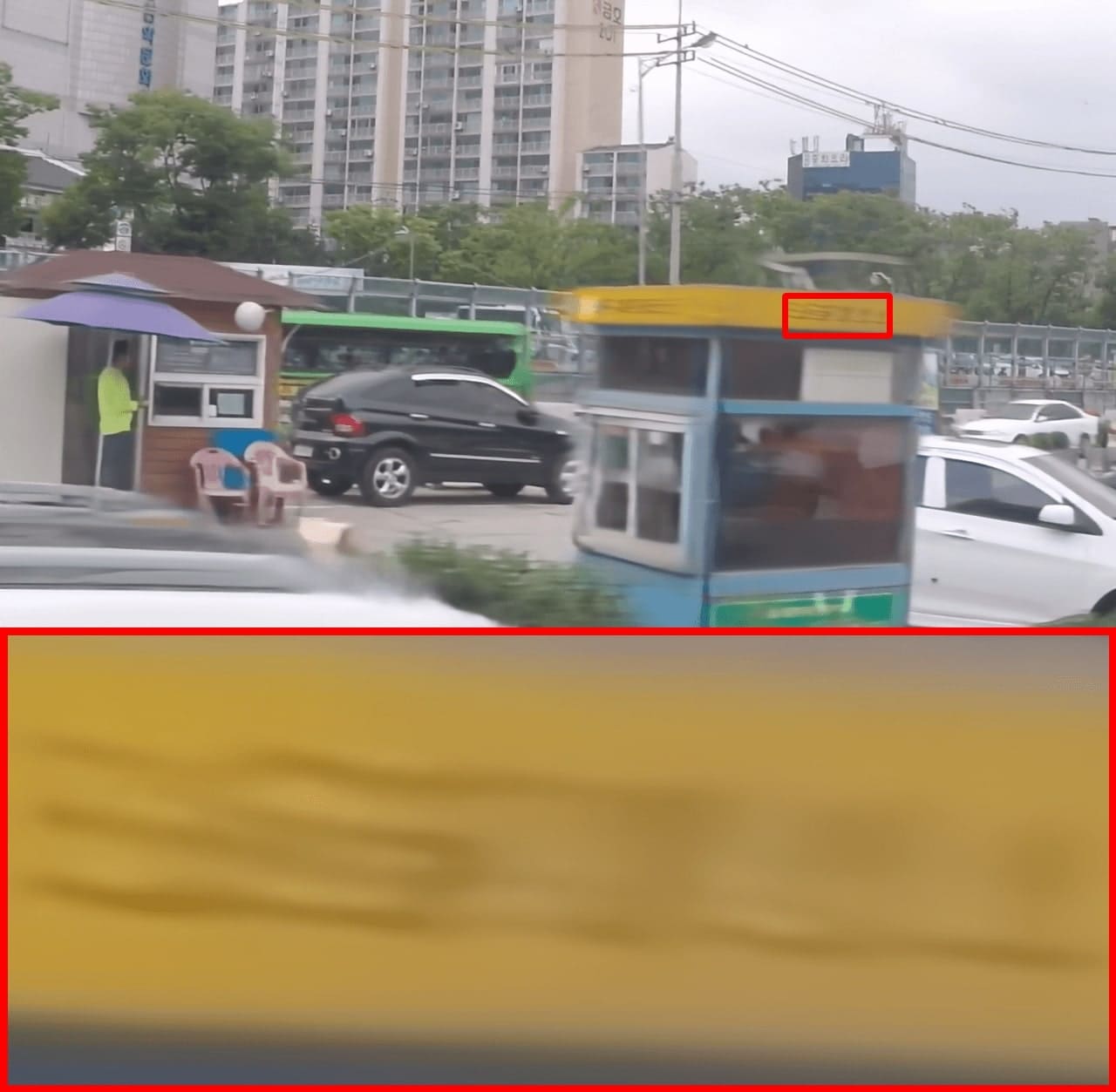}
        \subcaption{CDVD-TSP~\cite{pan2020cascaded}}
    \end{subfigure}
     \begin{subfigure}[t]{0.48\textwidth}
        \includegraphics[width=0.32\linewidth]{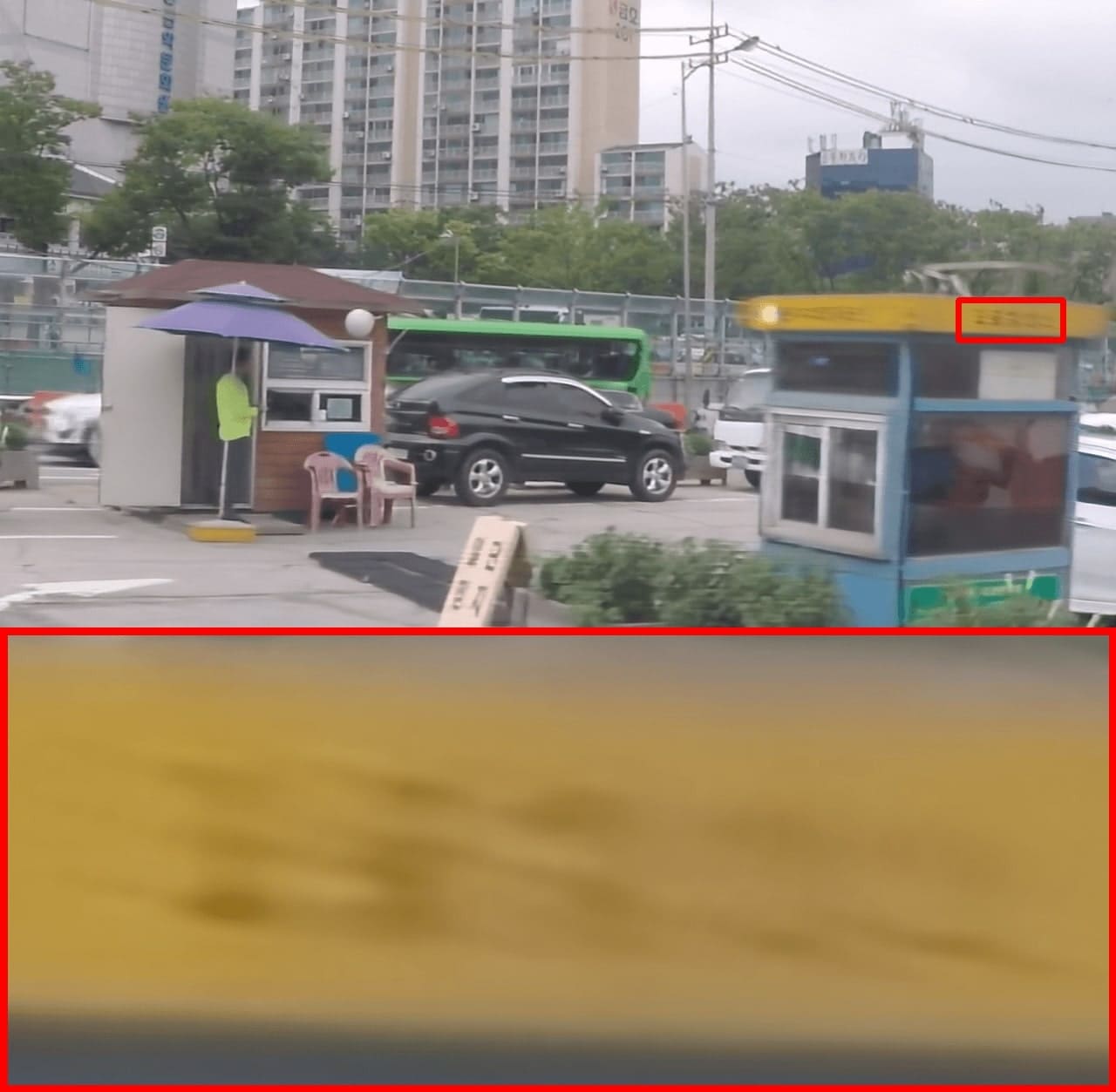}
        \includegraphics[width=0.32\linewidth]{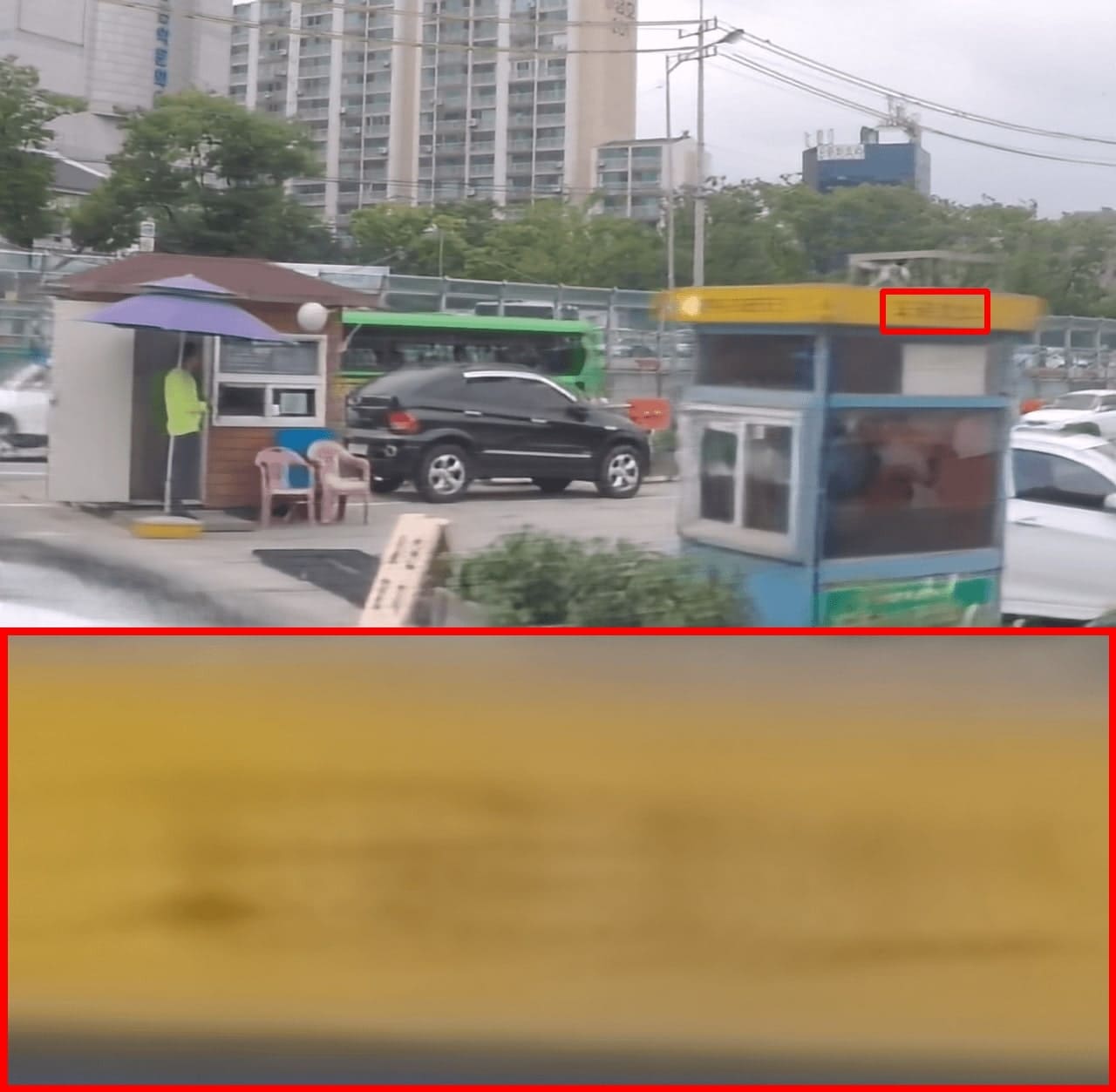}
        \includegraphics[width=0.32\linewidth]{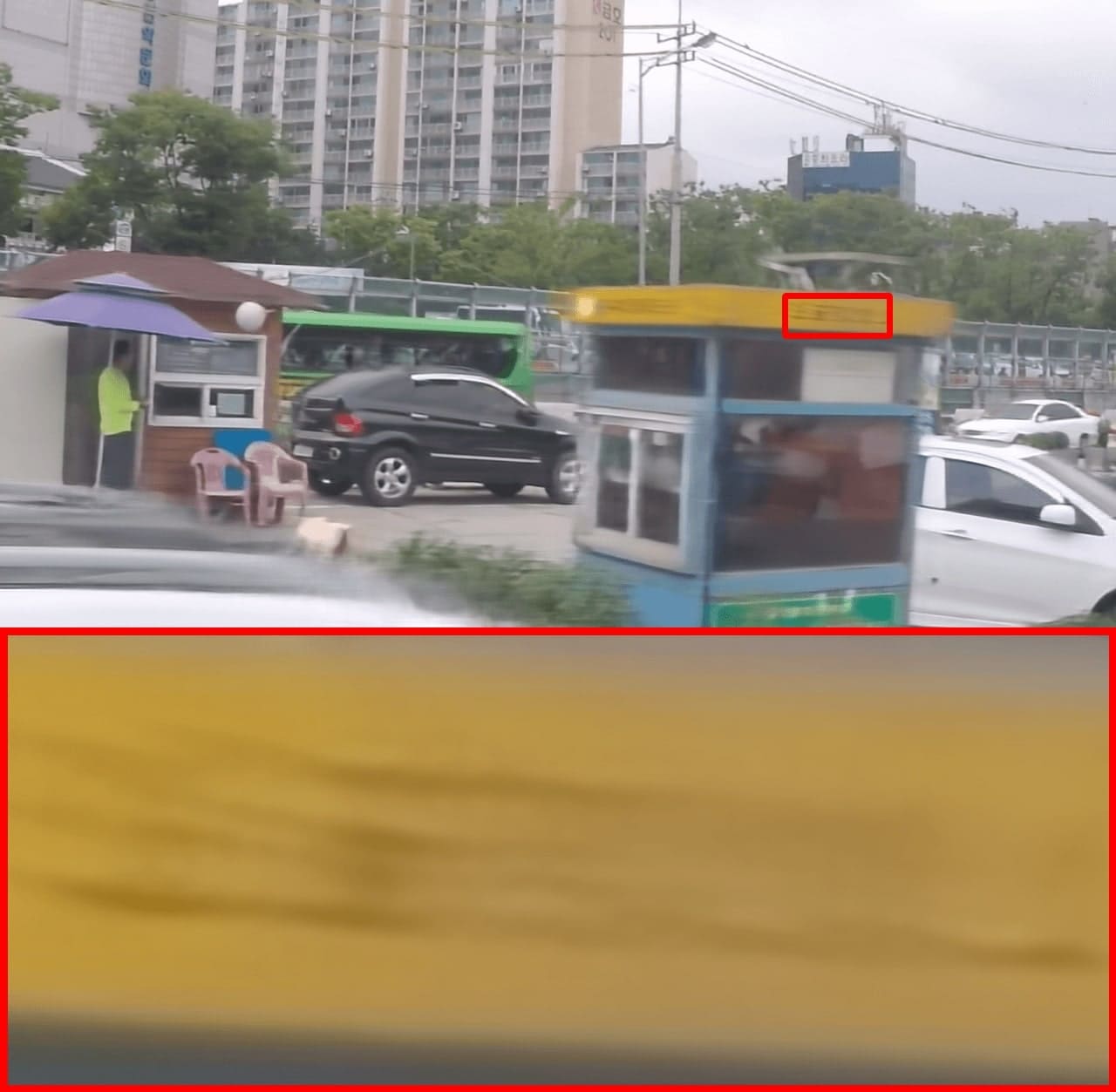}
        \subcaption{ESTRNN~\cite{zhong2020efficient}}
    \end{subfigure}   
    \begin{subfigure}[t]{0.48\textwidth}
        \includegraphics[width=0.32\linewidth]{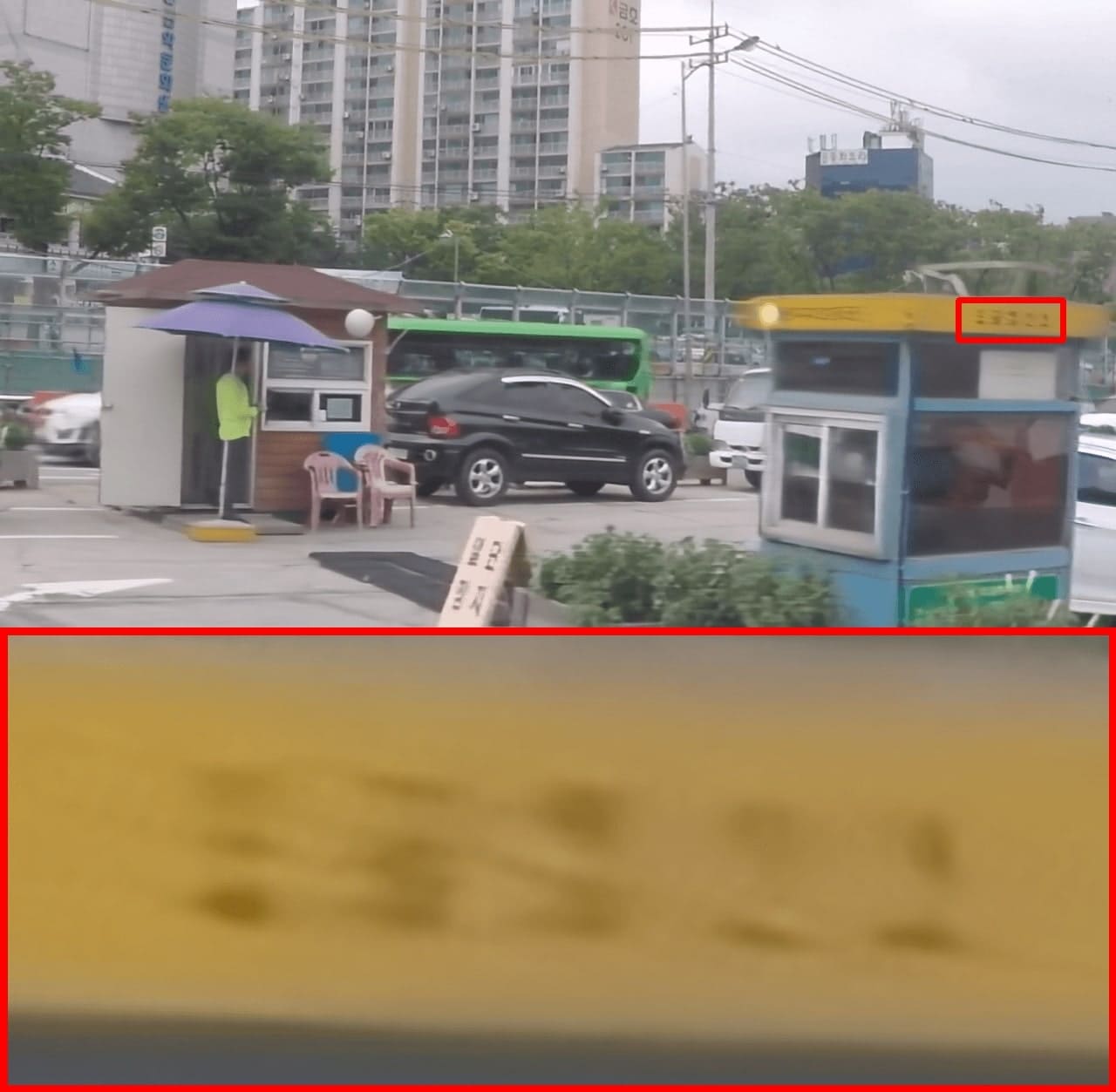}
        \includegraphics[width=0.32\linewidth]{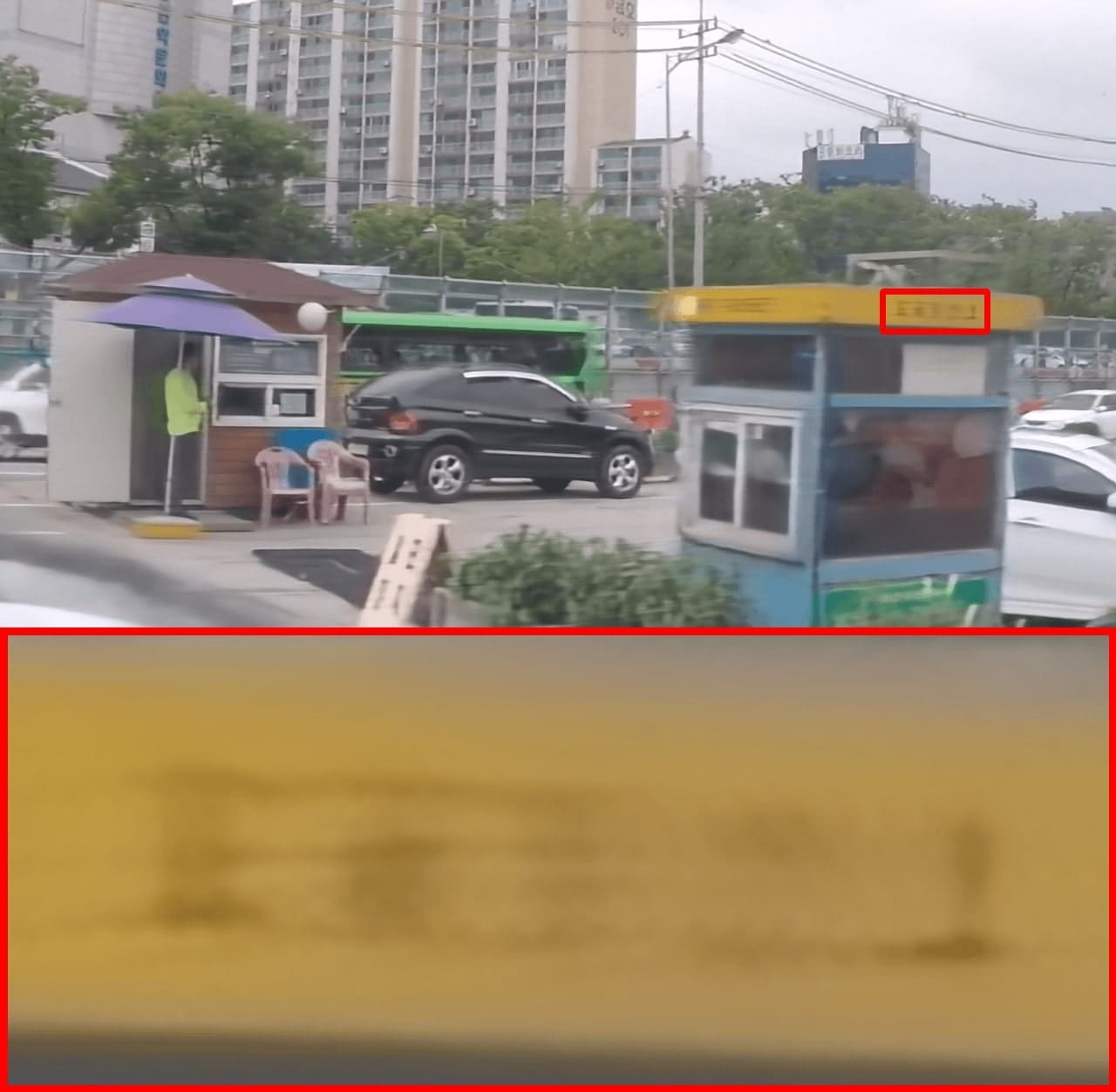}
        \includegraphics[width=0.32\linewidth]{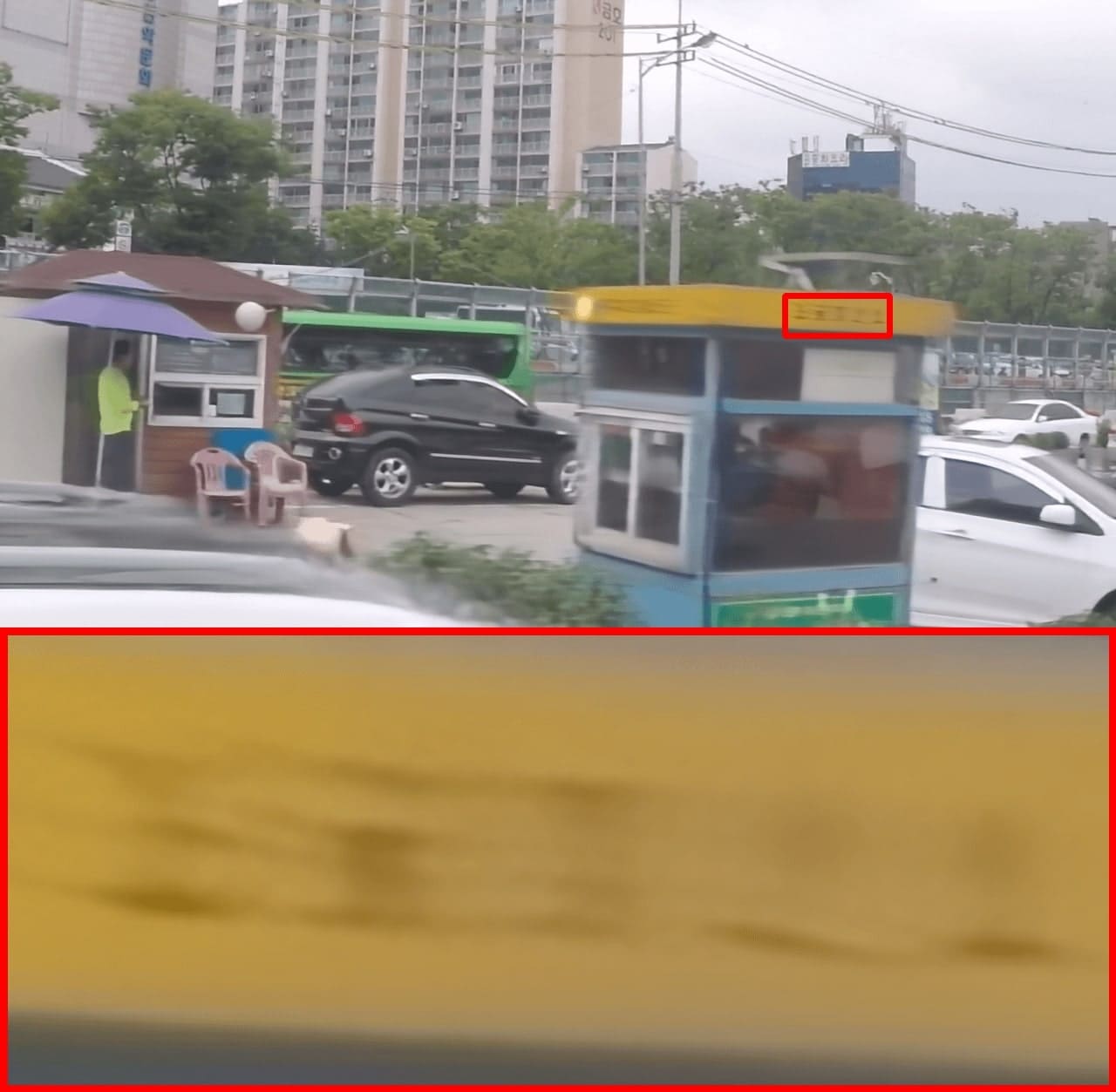}
        \subcaption{Ours}
    \end{subfigure}
    \begin{subfigure}[t]{0.48\textwidth}
        \includegraphics[width=0.32\linewidth]{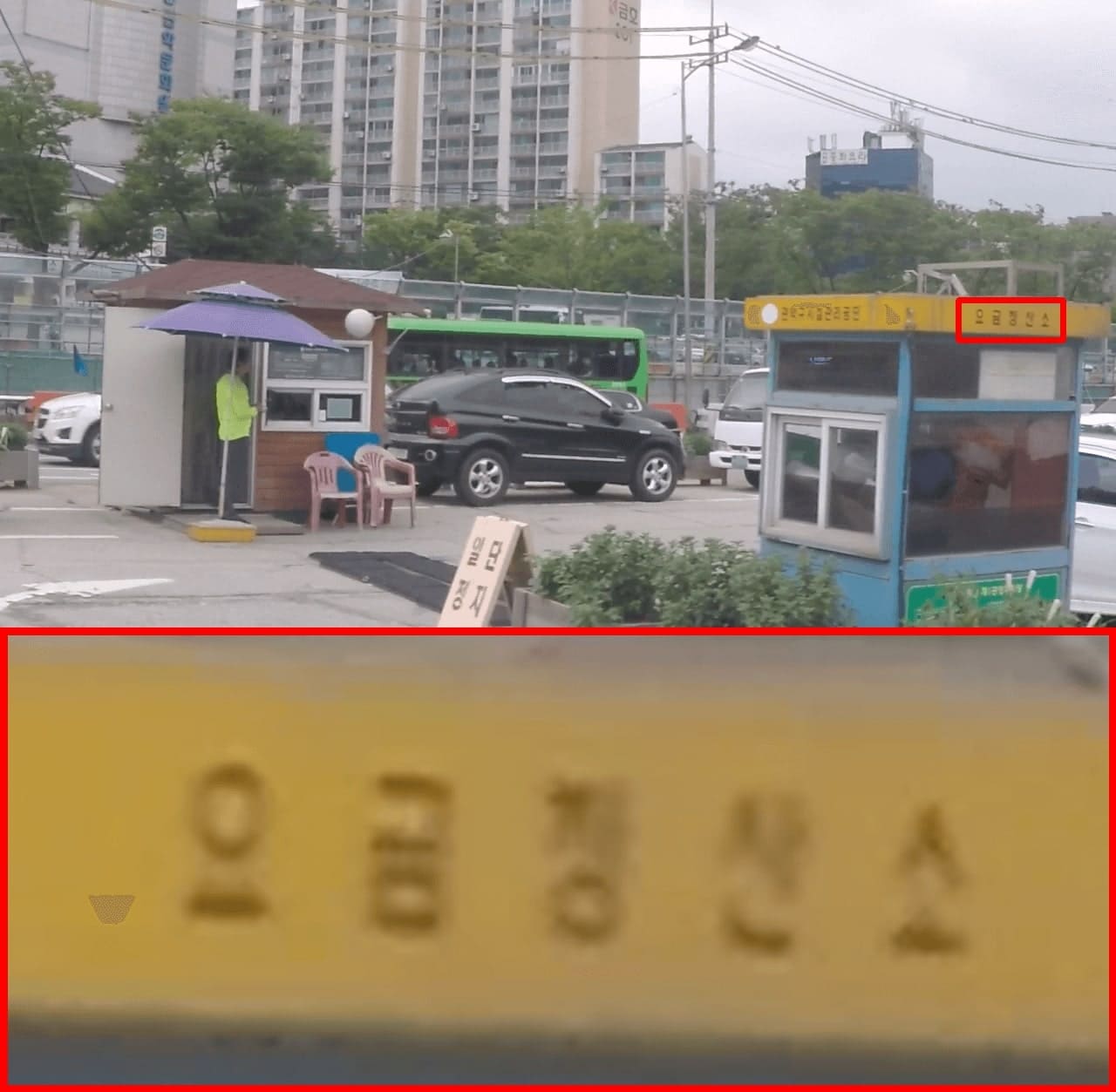}
        \includegraphics[width=0.32\linewidth]{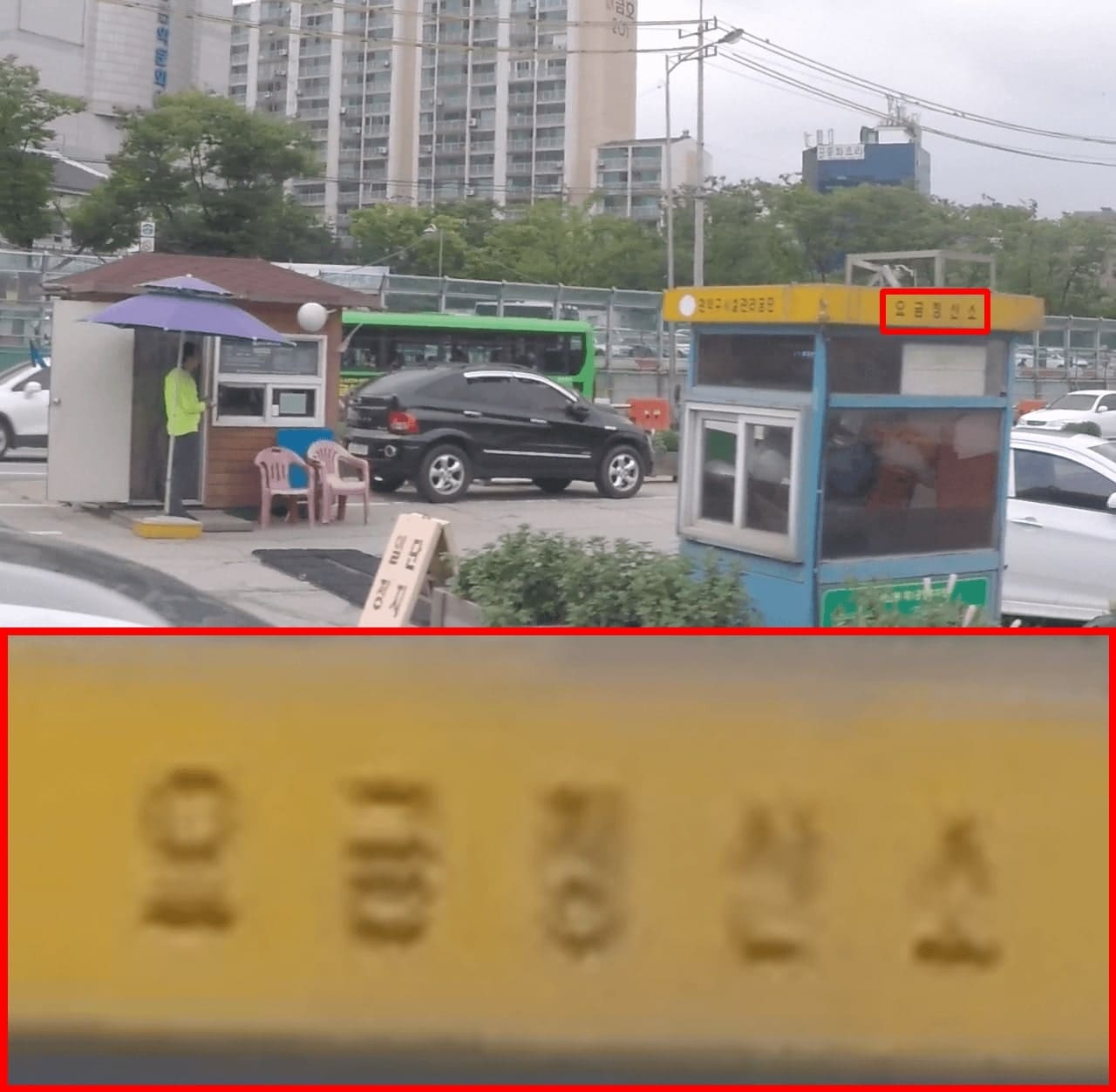}
        \includegraphics[width=0.32\linewidth]{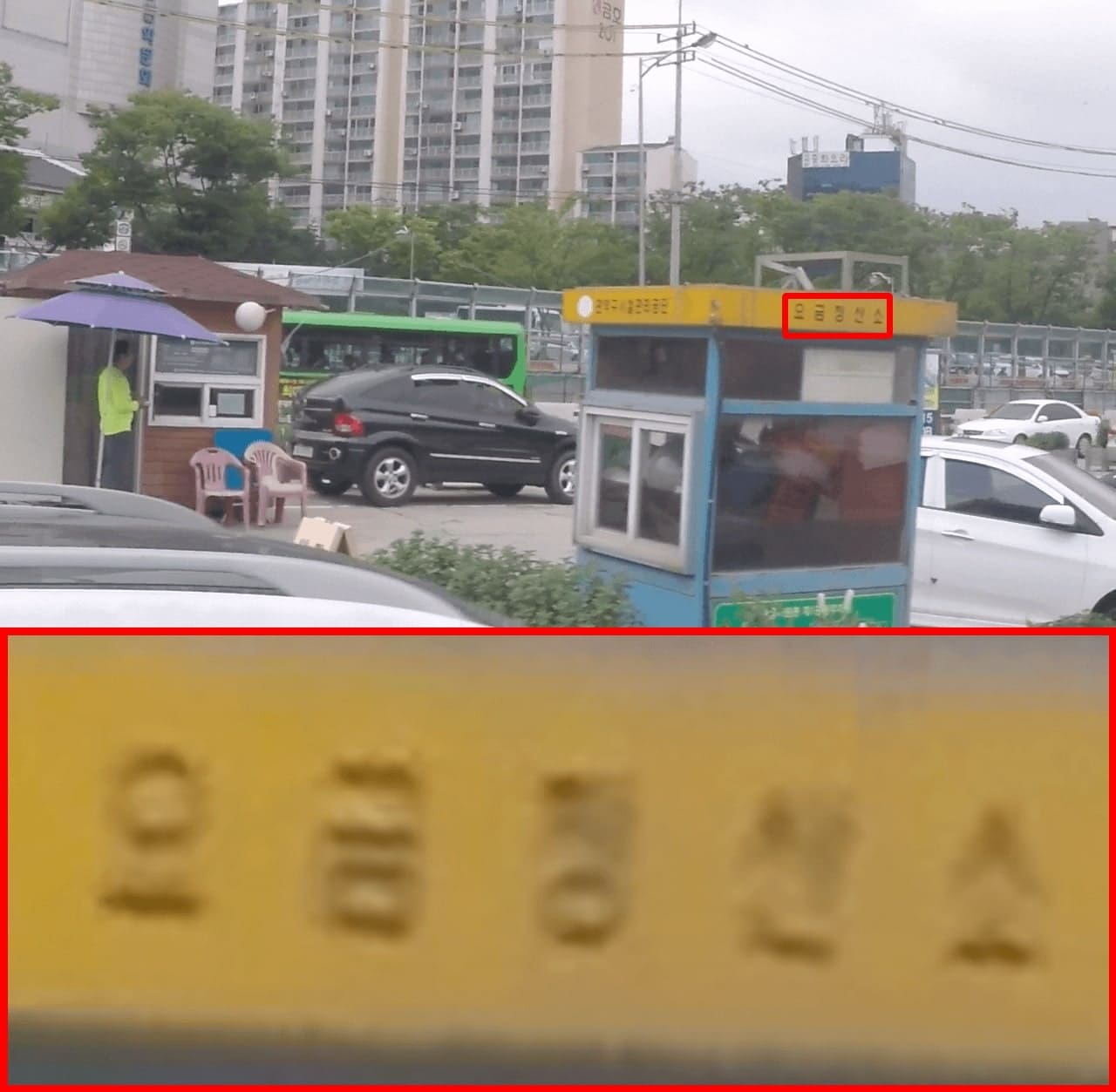}
        \subcaption{Ground-truth}
    \end{subfigure}
    \vspace{-0.1cm}
    \caption{Qualitative comparisons of frames with severe blur on the original GOPRO dataset~\cite{nah2017deep}.   \label{fig:qualitative_gopro2}}
    \vspace{-1em}
\end{figure*}

\begin{figure*}
    \centering
    \begin{subfigure}[t]{0.18\textwidth}
        \includegraphics[width=\linewidth]{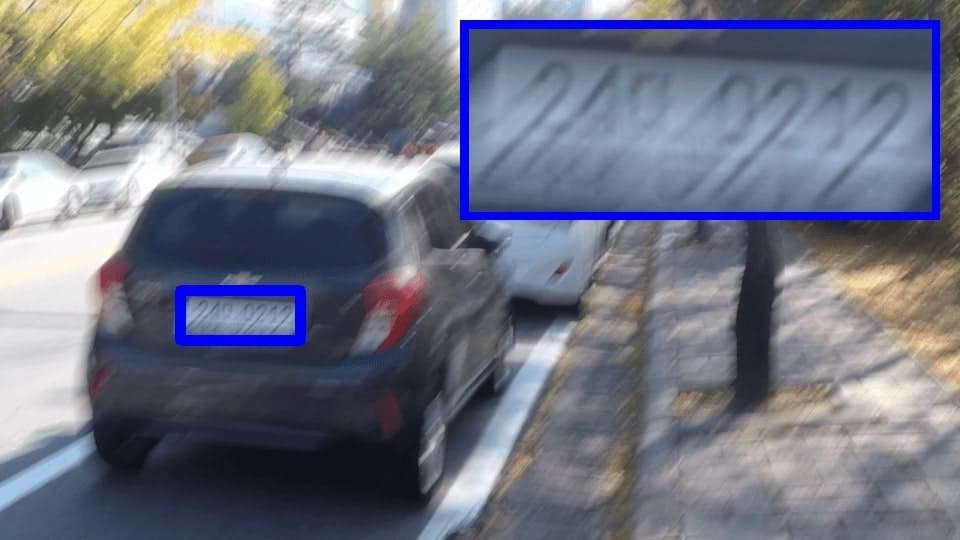}
        \subcaption{Query frame and pattern}\label{fig:query_frame_pattern}
    \end{subfigure}
    \begin{subfigure}[t]{0.19\textwidth}
        \includegraphics[width=\linewidth]{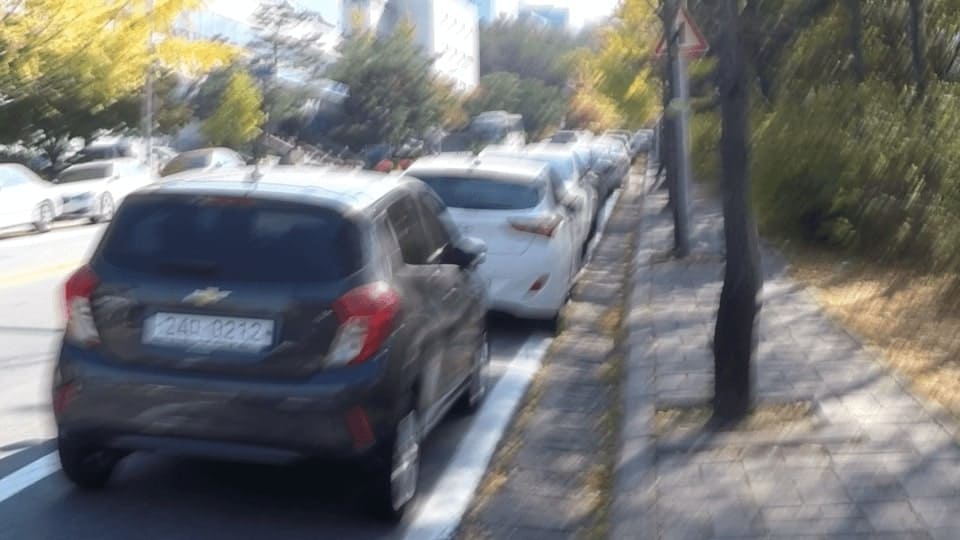}
        \subcaption{Matched frame}\label{fig:matched_frame}
    \end{subfigure}
    \begin{subfigure}[t]{0.19\textwidth}
        \includegraphics[width=\linewidth]{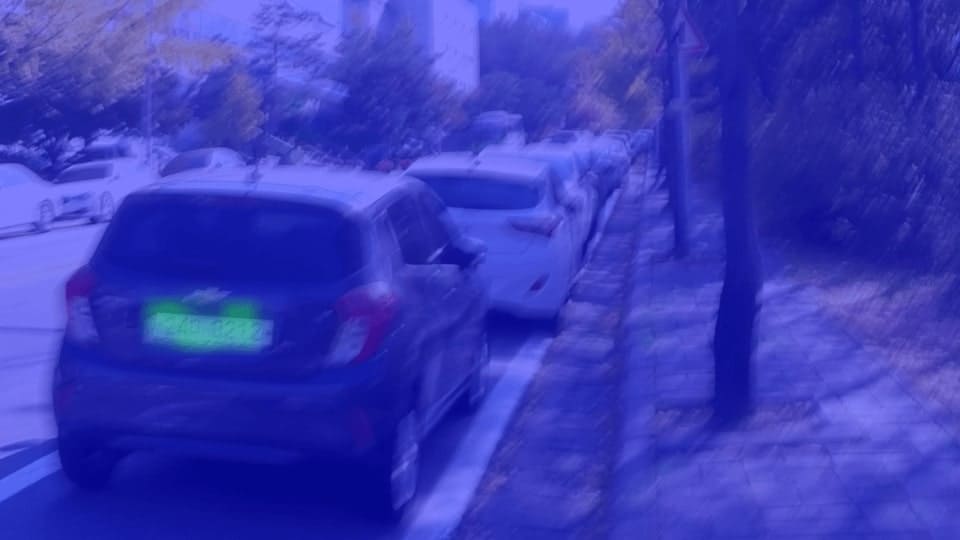}
        \subcaption{Attention map}\label{fig:attention_map}
    \end{subfigure}
    \begin{subfigure}[t]{0.19\textwidth}
        \includegraphics[width=\linewidth]{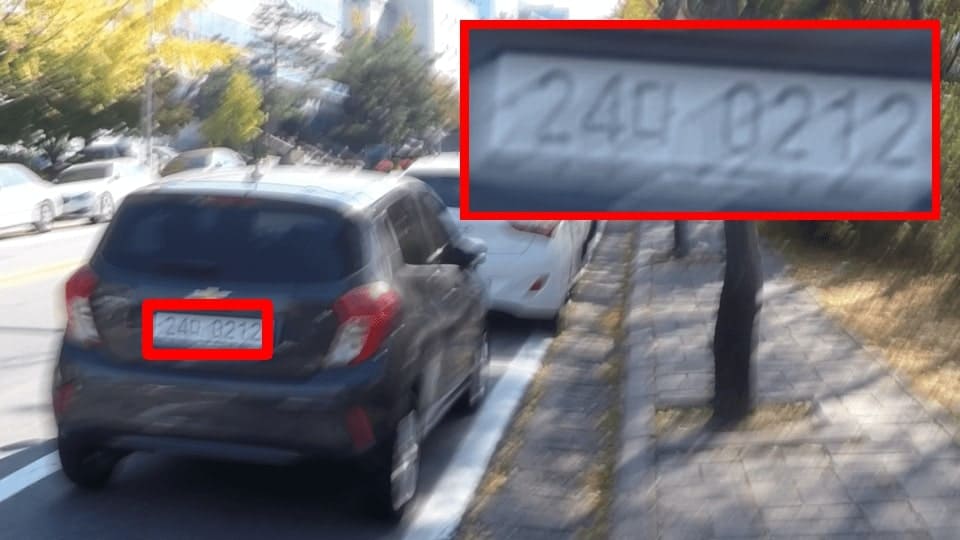}
        \subcaption{Matched pattern}\label{fig:matched_pattern}
    \end{subfigure}
    \begin{subfigure}[t]{0.19\textwidth}
        \includegraphics[width=\linewidth]{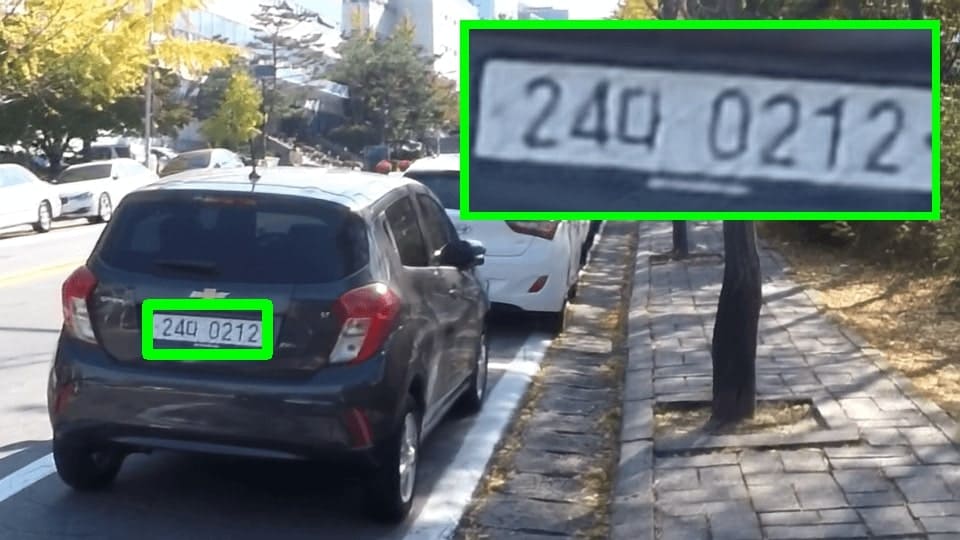}
        \subcaption{Matched deblurred result}\label{fig:sharp_frame_pattern}
    \end{subfigure}
    \vspace{-0.1cm}
    \caption{Visualization of spatio-temporal attention map obtained by our memory branch. \label{fig:vis_memory}}
    \vspace{-1em}
\end{figure*}

A quantitative comparison on the Downsampled GoPro dataset shown in Table~\ref{table:sota_downsampled_gopro} and on the Original GoPro dataset shown in Table~\ref{table:sota_original_gopro} 
indicates that our model achieves the best performance compared with other state-of-the-art models.  For Downsampled GoPro, we designed a slim version of our model to match our GMACs with ESTRNN~\cite{zhong2020efficient}. The slim version does not use the bidirectional and multi-scale design, yet it outperforms other state-of-the-art methods. Table~\ref{table:sota_runtime} also shows that our runtime is less than ESTRNN. 
Our final model, which includes the bidirectional and multi-scale structures, uses $147.15$ GMACs more than the slim version and improves performance by $0.57$dB. 

On the Original GoPro dataset in Table~\ref{table:sota_original_gopro}, the closest competing state-of-the-art model is CDVD-TSP~\cite{pan2020cascaded}.  Our model has $0.09$dB higher PSNR than CDVD-TSP even though the GMACs is an order of magnitude less (7\% specifically) and the runtime is one-fifth that of CDVD-TSP (see Table~\ref{table:sota_runtime}).  CDVD-TSP has significantly higher computational complexity, because it relies on optical flow and uses a cascaded training strategy.

We present the qualitative comparisons in Fig.~\ref{fig:qualitative_gopro} and ~\ref{fig:qualitative_gopro2}. Fig.~\ref{fig:qualitative_gopro} shows the performance on a frame with mild blur, while Fig.~\ref{fig:qualitative_gopro2} presents the performance on consecutive frames with severe blur due to object motion. Our model recovers clearer and sharper images than other state-of-the-art models. Even on video sequences with particularly large motion and severe blur, our method can still recover some image structures. For example, in  Fig.~\ref{fig:qualitative_gopro2}, only our deblurred images can be seen to have roughly $5$ characters. This is because our memory bank aggregates information of the whole video sequence and stores diverse features, so that when processing extremely blurred images, we can find relevant memory features from the memory bank to help restore them. More results are provided in the Supplementary.

\begin{figure}
    \centering
    \begin{subfigure}[t]{0.64\columnwidth}
        \includegraphics[width=\linewidth]{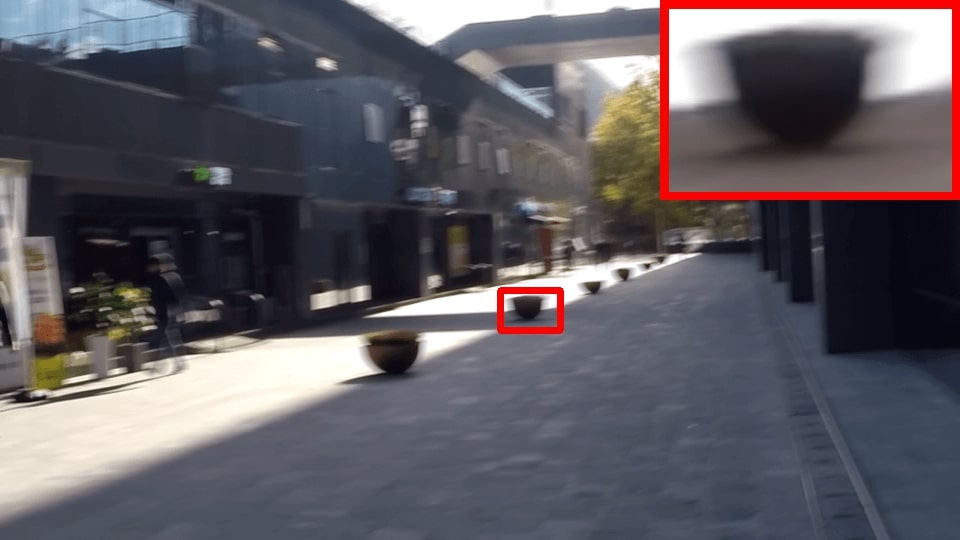}
        \subcaption{Query frame and pattern}\label{fig:vis_query}
    \end{subfigure}
    \begin{subfigure}[t]{0.32\columnwidth}
        \includegraphics[width=\linewidth]{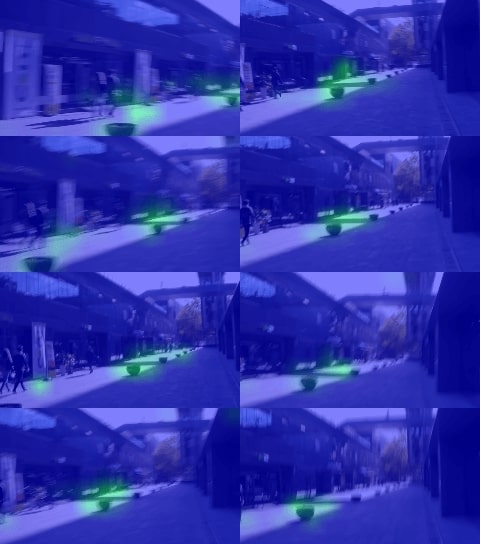}
        \subcaption{$s=3$}\label{fig:vis_s3}
    \end{subfigure}
    \begin{subfigure}[t]{0.64\columnwidth}
        \includegraphics[width=\linewidth]{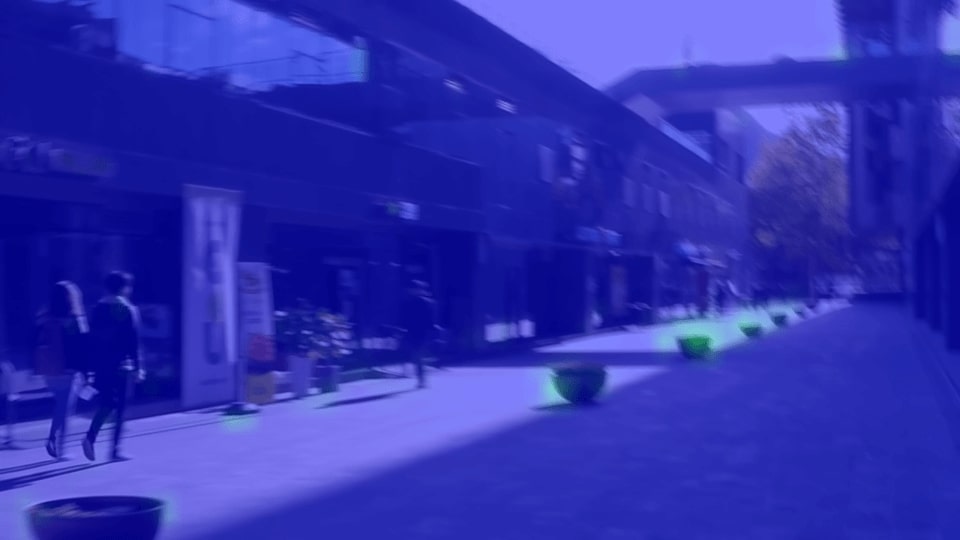}
        \subcaption{$s=1$}\label{fig:vs_s2}
    \end{subfigure}
    \begin{subfigure}[t]{0.32\columnwidth}
        \includegraphics[width=\linewidth]{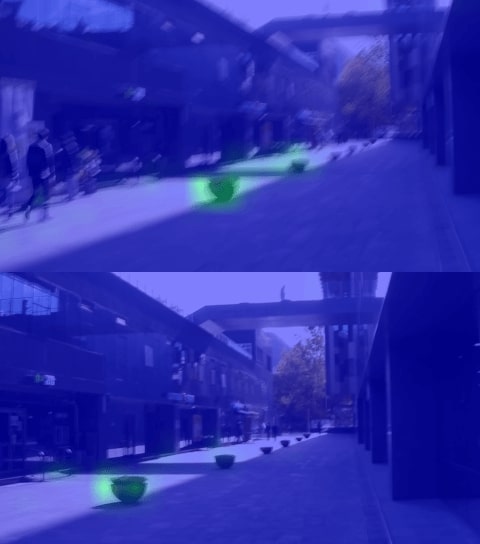}
        \subcaption{$s=2$}\label{fig:vis_s1}
    \end{subfigure}
    \vspace{-0.3cm}
    \caption{Visualization of attention map with multi-scale architecture. The same patterns that reoccur but are not of the same scale are matched. Note that all these matched memories contribute to the deblurring task for the input. Zoom in for a better view.}
    \label{fig:vis_memory_multi_scale}
    \vspace{-0.5em}
\end{figure}

\section{Network analysis}\label{sec:network_analysis}
We conducted extensive ablation studies to verify our approach. All ablations in this section were performed on Downsampled GOPRO~\cite{nah2017deep}.

\textbf{Effects of proposed components.} We evaluate 
the memory branch, bidirectional recurrency and our multi-scale strategy and show results in Table~\ref{table:ablation_study}. The vanilla version with deblurring branch only achieves the worst performance. Adding the proposed components increases both PSNR and SSIM.  The bidirectionality has slightly more impact than the multi-scale architecture, though the two components complement each other.

\begin{table}
    \caption{Ablation of model components.\label{table:ablation_study}}
    \vspace{-0.1cm}
    \centering
    \resizebox{\columnwidth}{!}{
    \def\arraystretch{1.1}
    \begin{tabular}{@{\extracolsep{4pt}}ccccc @{}}
        \Xhline{3\arrayrulewidth}
         Memory & Bidirection & Multi-scale & PSNR & SSIM\\
        \hline
        &  & & 30.88 & 0.9145 \\
         \checkmark &  & & 31.22 & 0.9203 \\
        \checkmark & \checkmark & & 31.63 & 0.9256 \\
        \checkmark &  & \checkmark & 31.44 & 0.9237 \\
        \checkmark & \checkmark & \checkmark & \textbf{31.79} & \textbf{0.9278} \\
        \Xhline{3\arrayrulewidth}
    \end{tabular}
    }
    \vspace{-1.5em}
\end{table}

\begin{figure}
    \centering
    \begin{subfigure}[t]{0.49\columnwidth}
        \includegraphics[width=\linewidth]{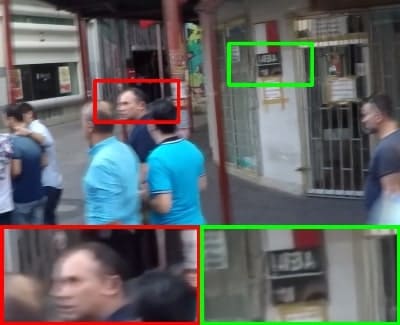}
        \subcaption{Naive memory network}
    \end{subfigure}
    \begin{subfigure}[t]{0.49\columnwidth}
        \includegraphics[width=\linewidth]{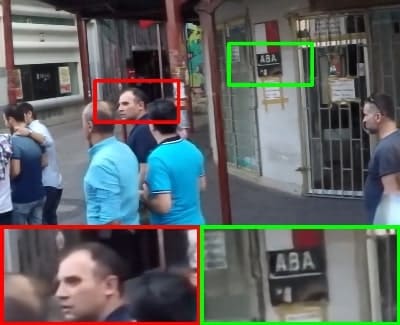}
        \subcaption{Our memory solution}
    \end{subfigure}
    \vspace{-0.1cm}
    \caption{Qualitative comparison between naive memory approach and our memory network. \label{fig:vis_memory_naive}}
    \vspace{-0.5cm}
\end{figure}

\textbf{Memory branch design.} 
We first followed a naive approach from the original memory network~\cite{oh2019video,weston2014memory,sukhbaatar2015end}, where we only had memory branches and decoded the memory directly into a sharp frame with an upsample module (\emph{`w/o Deblur.'} in Table~\ref{table:memory_branch}), to establish our baseline.
Fig.~\ref{fig:vis_memory_naive} shows that this naive form of stand-alone memory is insufficient to solve the deblurring task.  The quantitative measures in Table~\ref{table:memory_branch} support this finding. The results illustrate that for video deblurring, the deep network used to extract deep low-level features is essential.

We also experimented with a memory variant, which always stores the previous adjacent blurry-sharp feature pairs as a temporary item in the memory bank ~\cite{oh2019video} (\emph{`Temp'} in Table~\ref{table:memory_branch}). This approach increases the complexity but hurts the performance.  
This is because the adjacent frames are similar and contain fine details for sharpening the current frame.
Saving them in the memory space is likely to compress out these details and the corresponding attention weights will be distracted by other memories, as the attention is calculated on the entire memory bank.
This motivated us to design a more straightforward approach to directly concatenate the previously restored sharp results with the current frame
and feed them into the network. Since the motion between adjacent frames is generally small, our model can capture the correlation well. This approach achieves the best result as shown in Table~\ref{table:memory_branch}. 

\begin{table}
    \caption{Experimental results on high-level designs of our models. The `w/o Mem.' and `w/o Deblur.' columns show the results without memory branch and deblurring branch, respectively; the `Temp.' column shows the results using the temporary memory for the previous adjacent frame.  \label{table:memory_branch}}
    \vspace{-0.1cm}
    \centering
    \resizebox{\columnwidth}{!}{
    \def\arraystretch{1.1}
    \begin{tabular}{@{\extracolsep{4pt}}ccccc @{}}
        \Xhline{3\arrayrulewidth}
        Design & w/o Mem. & w/o Deblur. & Temp. & Ours \\
        \hline
        PSNR & 30.88 & 30.58 & 31.03 & \textbf{31.22} \\
        SSIM & 0.9145 & 0.9105 & 0.9169 & \textbf{0.9203}\\
        \Xhline{3\arrayrulewidth}
    \end{tabular}
    }
    \vspace{-1em}
\end{table}

\textbf{Separate $h$ and $r$ memories.} Previous methods using memory networks work well when encoding only the target output, \eg only the segmentation map in object segmentation~\cite{oh2019video,cheng2021rethinking}.  In our case, this would correspond to memorizing only the deblurred results $r$.  However, given the importance of the hidden feature $h$ in recurrent methods~\cite{nah2019recurrent,zhong2020efficient}, it is also logical to memorize $h$. Table~\ref{table:ablation_study_hr} shows that of these two, memorizing $h$ is more beneficial than $r$.  
It is also possible to concatenate the two (\emph{`$[h,r]$'}) and save them as a single memory value, but this gives only a marginal bump in performance compared to only memorizing $r$.  Storing the two in separate memories under the same key gives a much bigger boost.   

\begin{table}
    \caption{Comparison of different variants of utilizing features $h$ and $r$. $[h, r]$ means we concatenate two features and save them as a single memory value. \label{table:ablation_study_hr}}
    \vspace{-0.1cm}
    \centering
    \def\arraystretch{1.1}
    \begin{tabular}{@{\extracolsep{4pt}}ccccc @{}}
        \Xhline{3\arrayrulewidth}
        Variant & $h$ only & $r$ only & $[h, r]$ & Ours\\
        \hline
        PSNR & 31.03 & 30.94 & 31.00 & \textbf{31.22} \\
        SSIM & 0.9178 & 0.9152 & 0.9174 & \textbf{0.9203} \\
        \Xhline{3\arrayrulewidth}
    \end{tabular}
    \vspace{-1.5em}
\end{table}

\textbf{Memorization period.} As mentioned in Section~\ref{sec:memory}, we memorize every $T$ frames. As we adopt the multi-scale strategy, for scale levels larger than $1$, the input resolution is small, so decreasing $T$ 
does not result in a large amount of computation. However, for the scale level $s\!=\!1$, decreasing the memory period $T_1$
results in a significant increase in the computational and memory cost. We performed the experiment on scale level of $s\!=\!1$ to evaluate the effect of memorization period. Table~\ref{table:memorization_comparison} 
shows that even though memorizing every frame performs the best, its runtime is 
expensive. Therefore, to balance the trade-off between efficiency and performance, we memorize every $5$ frames at scale level $s\!=\!1$. Similarly, for $s\!=\!2$ and $s\!=\!3$, we empirically memorize every $2$ and $1$ frame, respectively. 

\begin{table}
    \caption{Memorization period comparison. The PSNR is calculated on the downsampled GOPRO dataset. \label{table:memorization_comparison}}
    \vspace{-0.1cm}
    \centering
    \def\arraystretch{1.1}
    \begin{tabular}{@{\extracolsep{4pt}}ccccc @{}}
        \Xhline{3\arrayrulewidth}
        T & PSNR & SSIM & GMACs & Runtime (s) \\
        \hline
        1 & \textbf{31.42} & \textbf{0.9231} & 218.79 & 0.582 \\
        3 & 31.21 & 0.9202 & 201.10 & 0.154 \\
        5 & 31.22 & 0.9203 & \textbf{197.34} & \textbf{0.095} \\       \Xhline{3\arrayrulewidth}
    \end{tabular}
\vspace{-1.5em}
\end{table}

\textbf{Visualizations.}
To get a better sense of what the memory retrieves for a query pattern, we calculated the attention map encoded in $\mathbf{W}$ in Eq.~\ref{eq:calc_w} on the memory bank.
The visualized result is shown in Fig.~\ref{fig:vis_memory}, where we select one of the matched frames for reference.
It can be observed that since we need to deblur the license plate number, most of the attention is given to the license plate number in the matched frame.
The attention maps for the multi-scale memory are shown in Fig.~\ref{fig:vis_memory_multi_scale}. We observe that the memory bank provides different scales of matching. This further enhances the query and utilization of features.

\textbf{Limitations.}
The size of the memory in the memory bank affects the GPU memory overhead and computational cost. In Section~\ref{sec:exp_setting}, we consider a simple  way of reducing the memory size by discarding old memories under the assumption that the most recent memories are more important.  We find this approach effective, but we believe that more principled memory management strategies can be considered for future work.

\section{Conclusion}

We proposed a multi-scale memory-based network for deep video deblurring, which saves blurry-sharp feature pairs in the memory bank. To restore a blurry input frame, we retrieve relevant information for each image region by performing the spatio-temporal attention in the memory bank. The memory bank allows region-aware information retrieval to achieve fine-grained deblurring. To enrich the diversity and utility of the memory bank, we developed a bidirectional and multi-scale strategy. Both quantitative and qualitative experimental results show that our proposal outperforms the state-of-the-art models while maintaining a minimal cost in computational complexity and runtime.

\section{Acknowledgement}
This research is supported by the National Research Foundation, Singapore under its NRF Fellowship for AI (NRF-NRFFAI1-2019-0001). Any opinions, findings and conclusions or recommendations expressed in this material are those of the author(s) and do not reflect the views of National Research Foundation, Singapore.

{\small
\bibliographystyle{ieee_fullname}
\bibliography{PaperForArxiv}

\begin{thebibliography}{10}\itemsep=-1pt

\bibitem{charbonnier1994two}
Pierre Charbonnier, Laure Blanc-Feraud, Gilles Aubert, and Michel Barlaud.
\newblock Two deterministic half-quadratic regularization algorithms for
  computed imaging.
\newblock In {\em Proceedings of 1st International Conference on Image
  Processing}, volume~2, pages 168--172. IEEE, 1994.

\bibitem{cheng2021rethinking}
Ho~Kei Cheng, Yu-Wing Tai, and Chi-Keung Tang.
\newblock Rethinking space-time networks with improved memory coverage for
  efficient video object segmentation.
\newblock {\em arXiv preprint arXiv:2106.05210}, 2021.

\bibitem{gast2019deep}
Jochen Gast and Stefan Roth.
\newblock Deep video deblurring: The devil is in the details.
\newblock In {\em Proceedings of the IEEE/CVF International Conference on
  Computer Vision Workshops}, pages 0--0, 2019.

\bibitem{guo2019toward}
Shi Guo, Zifei Yan, Kai Zhang, Wangmeng Zuo, and Lei Zhang.
\newblock Toward convolutional blind denoising of real photographs.
\newblock In {\em Proceedings of the IEEE/CVF Conference on Computer Vision and
  Pattern Recognition}, pages 1712--1722, 2019.

\bibitem{he2016deep}
Kaiming He, Xiangyu Zhang, Shaoqing Ren, and Jian Sun.
\newblock Deep residual learning for image recognition.
\newblock In {\em Proceedings of the IEEE conference on computer vision and
  pattern recognition}, pages 770--778, 2016.

\bibitem{hyun2017online}
Tae Hyun~Kim, Kyoung Mu~Lee, Bernhard Scholkopf, and Michael Hirsch.
\newblock Online video deblurring via dynamic temporal blending network.
\newblock In {\em Proceedings of the IEEE International Conference on Computer
  Vision}, pages 4038--4047, 2017.

\bibitem{kim2020transfer}
Yoonsik Kim, Jae~Woong Soh, Gu~Yong Park, and Nam~Ik Cho.
\newblock Transfer learning from synthetic to real-noise denoising with
  adaptive instance normalization.
\newblock In {\em Proceedings of the IEEE/CVF Conference on Computer Vision and
  Pattern Recognition}, pages 3482--3492, 2020.

\bibitem{kingma2014adam}
Diederik~P Kingma and Jimmy Ba.
\newblock Adam: A method for stochastic optimization.
\newblock {\em arXiv preprint arXiv:1412.6980}, 2014.

\bibitem{lee2011simultaneous}
Hee~Seok Lee, Junghyun Kwon, and Kyoung~Mu Lee.
\newblock Simultaneous localization, mapping and deblurring.
\newblock In {\em 2011 International Conference on Computer Vision}, pages
  1203--1210. IEEE, 2011.

\bibitem{lim2017enhanced}
Bee Lim, Sanghyun Son, Heewon Kim, Seungjun Nah, and Kyoung Mu~Lee.
\newblock Enhanced deep residual networks for single image super-resolution.
\newblock In {\em Proceedings of the IEEE conference on computer vision and
  pattern recognition workshops}, pages 136--144, 2017.

\bibitem{lu2020video}
Xiankai Lu, Wenguan Wang, Martin Danelljan, Tianfei Zhou, Jianbing Shen, and
  Luc Van~Gool.
\newblock Video object segmentation with episodic graph memory networks.
\newblock In {\em Computer Vision--ECCV 2020: 16th European Conference,
  Glasgow, UK, August 23--28, 2020, Proceedings, Part III 16}, pages 661--679.
  Springer, 2020.

\bibitem{miller2016key}
Alexander Miller, Adam Fisch, Jesse Dodge, Amir-Hossein Karimi, Antoine Bordes,
  and Jason Weston.
\newblock Key-value memory networks for directly reading documents.
\newblock {\em arXiv preprint arXiv:1606.03126}, 2016.

\bibitem{na2017read}
Seil Na, Sangho Lee, Jisung Kim, and Gunhee Kim.
\newblock A read-write memory network for movie story understanding.
\newblock In {\em Proceedings of the IEEE International Conference on Computer
  Vision}, pages 677--685, 2017.

\bibitem{nah2017deep}
Seungjun Nah, Tae Hyun~Kim, and Kyoung Mu~Lee.
\newblock Deep multi-scale convolutional neural network for dynamic scene
  deblurring.
\newblock In {\em Proceedings of the IEEE conference on computer vision and
  pattern recognition}, pages 3883--3891, 2017.

\bibitem{nah2019recurrent}
Seungjun Nah, Sanghyun Son, and Kyoung~Mu Lee.
\newblock Recurrent neural networks with intra-frame iterations for video
  deblurring.
\newblock In {\em Proceedings of the IEEE/CVF Conference on Computer Vision and
  Pattern Recognition}, pages 8102--8111, 2019.

\bibitem{oh2019video}
Seoung~Wug Oh, Joon-Young Lee, Ning Xu, and Seon~Joo Kim.
\newblock Video object segmentation using space-time memory networks.
\newblock In {\em Proceedings of the IEEE/CVF International Conference on
  Computer Vision}, pages 9226--9235, 2019.

\bibitem{pan2020cascaded}
Jinshan Pan, Haoran Bai, and Jinhui Tang.
\newblock Cascaded deep video deblurring using temporal sharpness prior.
\newblock In {\em Proceedings of the IEEE/CVF Conference on Computer Vision and
  Pattern Recognition}, pages 3043--3051, 2020.

\bibitem{sajjadi2018frame}
Mehdi~SM Sajjadi, Raviteja Vemulapalli, and Matthew Brown.
\newblock Frame-recurrent video super-resolution.
\newblock In {\em Proceedings of the IEEE Conference on Computer Vision and
  Pattern Recognition}, pages 6626--6634, 2018.

\bibitem{seok2013dense}
Hee Seok~Lee and Kuoung Mu~Lee.
\newblock Dense 3d reconstruction from severely blurred images using a single
  moving camera.
\newblock In {\em Proceedings of the IEEE Conference on Computer Vision and
  Pattern Recognition}, pages 273--280, 2013.

\bibitem{seong2020kernelized}
Hongje Seong, Junhyuk Hyun, and Euntai Kim.
\newblock Kernelized memory network for video object segmentation.
\newblock In {\em European Conference on Computer Vision}, pages 629--645.
  Springer, 2020.

\bibitem{shi2016real}
Wenzhe Shi, Jose Caballero, Ferenc Husz{\'a}r, Johannes Totz, Andrew~P Aitken,
  Rob Bishop, Daniel Rueckert, and Zehan Wang.
\newblock Real-time single image and video super-resolution using an efficient
  sub-pixel convolutional neural network.
\newblock In {\em Proceedings of the IEEE conference on computer vision and
  pattern recognition}, pages 1874--1883, 2016.

\bibitem{su2017deep}
Shuochen Su, Mauricio Delbracio, Jue Wang, Guillermo Sapiro, Wolfgang Heidrich,
  and Oliver Wang.
\newblock Deep video deblurring for hand-held cameras.
\newblock In {\em Proceedings of the IEEE Conference on Computer Vision and
  Pattern Recognition}, pages 1279--1288, 2017.

\bibitem{suin2021gated}
Maitreya Suin and AN Rajagopalan.
\newblock Gated spatio-temporal attention-guided video deblurring.
\newblock In {\em Proceedings of the IEEE/CVF Conference on Computer Vision and
  Pattern Recognition}, pages 7802--7811, 2021.

\bibitem{sukhbaatar2015end}
Sainbayar Sukhbaatar, Arthur Szlam, Jason Weston, and Rob Fergus.
\newblock End-to-end memory networks.
\newblock {\em arXiv preprint arXiv:1503.08895}, 2015.

\bibitem{tao2018scale}
Xin Tao, Hongyun Gao, Xiaoyong Shen, Jue Wang, and Jiaya Jia.
\newblock Scale-recurrent network for deep image deblurring.
\newblock In {\em Proceedings of the IEEE Conference on Computer Vision and
  Pattern Recognition}, pages 8174--8182, 2018.

\bibitem{vaswani2017attention}
Ashish Vaswani, Noam Shazeer, Niki Parmar, Jakob Uszkoreit, Llion Jones,
  Aidan~N Gomez, {\L}ukasz Kaiser, and Illia Polosukhin.
\newblock Attention is all you need.
\newblock In {\em Advances in neural information processing systems}, pages
  5998--6008, 2017.

\bibitem{wang2019edvr}
Xintao Wang, Kelvin~CK Chan, Ke Yu, Chao Dong, and Chen Change~Loy.
\newblock Edvr: Video restoration with enhanced deformable convolutional
  networks.
\newblock In {\em Proceedings of the IEEE/CVF Conference on Computer Vision and
  Pattern Recognition Workshops}, pages 0--0, 2019.

\bibitem{wang2004image}
Zhou Wang, Alan~C Bovik, Hamid~R Sheikh, and Eero~P Simoncelli.
\newblock Image quality assessment: from error visibility to structural
  similarity.
\newblock {\em IEEE transactions on image processing}, 13(4):600--612, 2004.

\bibitem{weston2014memory}
Jason Weston, Sumit Chopra, and Antoine Bordes.
\newblock Memory networks.
\newblock {\em arXiv preprint arXiv:1410.3916}, 2014.

\bibitem{wieschollek2017learning}
Patrick Wieschollek, Michael Hirsch, Bernhard Scholkopf, and Hendrik Lensch.
\newblock Learning blind motion deblurring.
\newblock In {\em Proceedings of the IEEE International Conference on Computer
  Vision}, pages 231--240, 2017.

\bibitem{wu2011blurred}
Yi Wu, Haibin Ling, Jingyi Yu, Feng Li, Xue Mei, and Erkang Cheng.
\newblock Blurred target tracking by blur-driven tracker.
\newblock In {\em 2011 International Conference on Computer Vision}, pages
  1100--1107. IEEE, 2011.

\bibitem{yang2018learning}
Tianyu Yang and Antoni~B Chan.
\newblock Learning dynamic memory networks for object tracking.
\newblock In {\em Proceedings of the European conference on computer vision
  (ECCV)}, pages 152--167, 2018.

\bibitem{zhang2018image}
Yulun Zhang, Kunpeng Li, Kai Li, Lichen Wang, Bineng Zhong, and Yun Fu.
\newblock Image super-resolution using very deep residual channel attention
  networks.
\newblock In {\em Proceedings of the European conference on computer vision
  (ECCV)}, pages 286--301, 2018.

\bibitem{zhang2018residual}
Yulun Zhang, Yapeng Tian, Yu Kong, Bineng Zhong, and Yun Fu.
\newblock Residual dense network for image super-resolution.
\newblock In {\em Proceedings of the IEEE conference on computer vision and
  pattern recognition}, pages 2472--2481, 2018.

\bibitem{zhong2020efficient}
Zhihang Zhong, Ye Gao, Yinqiang Zheng, and Bo Zheng.
\newblock Efficient spatio-temporal recurrent neural network for video
  deblurring.
\newblock In {\em European Conference on Computer Vision}, pages 191--207.
  Springer, 2020.

\bibitem{zhou2019spatio}
Shangchen Zhou, Jiawei Zhang, Jinshan Pan, Haozhe Xie, Wangmeng Zuo, and Jimmy
  Ren.
\newblock Spatio-temporal filter adaptive network for video deblurring.
\newblock In {\em Proceedings of the IEEE/CVF International Conference on
  Computer Vision}, pages 2482--2491, 2019.

\end{thebibliography}
}

\end{document}